\documentclass[lettersize,journal]{IEEEtran}
\usepackage{subfigmat}
\usepackage{amsmath,amsfonts}
\usepackage{mathtools}
\usepackage{algorithm}
\usepackage{array}
\usepackage[caption=false,font=normalsize,labelfont=sf,textfont=sf]{subfig}
\newsubfloat{table}   %
\usepackage{textcomp}
\usepackage{stfloats}
\usepackage[hyphens]{url}
\usepackage{verbatim}
\usepackage{graphicx}
\usepackage{cite}
\usepackage{multirow}
\usepackage{multicol}
\usepackage{epstopdf}
\usepackage{mathcomp}
\usepackage{algpseudocode}
\usepackage{cleveref}
\usepackage{multirow}
\usepackage{lscape}
\usepackage{color}
\usepackage{xcolor}
\usepackage{adjustbox}

\newcommand{\argmin}{\mathop{\rm arg~min}\limits}

\newcommand{\vpanel}[2]{\subfigure[]{\adjustbox{margin=1pt,bgcolor=#1}{\includegraphics[width=3.4cm]{#2}}}}

\begin{document}

\title{Unrolled Time-Varying Graph Signal Restoration under Spatiotemporal Smoothness Priors}

\author{Hayate Kojima, Hikari Noguchi, \IEEEmembership{Graduate Student Member, IEEE,}
Koki Yamada, \IEEEmembership{Member, IEEE,}\\
Yuichi Tanaka, \IEEEmembership{Senior Member, IEEE}

\thanks{H. Kojima and Y. Tanaka are with the Graduate School of Engineering, The University of Osaka, Osaka, Japan. H. Noguchi and K. Yamada are with the Department of Electrical Engineering and Computer Science, Tokyo University of Agriculture and Technology, Koganei, Tokyo 184--8588, Japan (email: h-kojima@msp-lab.org; hnoguchi@msp-lab.org; k-yamada@go.tuat.ac.jp; ytanaka@comm.eng.osaka-u.ac.jp).
Our preliminary results are presented in \cite{kojima2023Restoration}. This work is supported in part by JSPS KAKENHI under Grant 25KJ1728, 24K20788, 23H01415, 23K26110, and 23K17461.}
}

\markboth{IEEE Transactions on Signal and Information Processing Over Networks,~Vol.~xx, 2026}%
{Kojima \MakeLowercase{\textit{et al.}}: Unrolled Time-Varying Graph Signal Restoration under Spatiotemporal Smoothness Priors}

\maketitle

\begin{abstract}
In this paper, we propose restoration methods for time-varying graph signals using deep algorithm unrolling (DAU). 
Time-varying graph signals, such as signals obtained from sensor networks, are nonuniformly distributed in space and observed as time series. 
Since these observed signals often contain noise and missing values, their restoration needs to consider both spatial and temporal relationships. 
Our approach is based on an optimization problem that models signal properties using a spatiotemporal regularizer that combines a Sobolev operator for spatial smoothness and a multi-tap FIR filter for temporal smoothness.
We then unroll the iterative conjugate gradient method to solve this problem and learn the regularization parameters and filter coefficients in each iteration.
Our method can be applied to supervised batch, supervised online, and unsupervised batch settings to learn these parameters.
Experiments on several synthetic and real-world datasets show that the supervised batch method achieves the lowest RMSEs in almost all cases.
The supervised online and the unsupervised batch methods outperform the existing methods of their own settings, and are often competitive with the existing batch methods.
\end{abstract}


\section{Introduction}
In many application fields such as wireless sensor networks, traffic monitoring, and environmental sensing, sensors are irregularly distributed over space.
Classical signal processing methods assume signals defined on regular equispaced domains, which limits their applicability to such irregularly structured data. 

Graph signal processing (GSP) extends signal processing tools to signals defined on graphs, where the graph models relationships among data samples \cite{shuman2013emerging,ortega2018Graph,leus2023Graph}. 
Over the past decade, GSP has attracted significant attention and has enabled a variety of tasks for graph signals, including sampling \cite{tanaka2020Sampling, hara2023Graph, yanagiya2025Edge, yamashita2025Controlling, wang2023Fast, peng2026RiskAware}, restoration \cite{6156471, qiu2017TimeVarying, giraldo2022Reconstruction, Graph_Signal_Restoration:Nagahama:TSP:2022, smola2003Kernels, rey2023Robust}, matrix completion \cite{buciulea2019Network, bagheri2024Joint}, filtering \cite{onuki2016Graph, ramakrishna2020User, yanagiya2024Lossy, fukuhara2025Period}, and graph learning \cite{xia2021Graph, yokota2026TimeVarying, yokota2025Efficient}.

In practice, graph signals are often corrupted by noise and missing observations due to sensing failures, communication constraints, or cost limitations. 
Restoring signals from such incomplete and noisy observations is therefore a fundamental task in GSP. 

Many early studies have focused on static graph signals, which represent a signal at a single time instance. 
These methods typically assume spatial smoothness of the signal on the graph and solve optimization problems using prior(s) such as Laplacian or Tikhonov regularization \cite{smola2003Kernels, perraudin2017Stationary}. 
However, many real-world graph signals evolve over time. 
Examples include traffic flows, environmental measurements, and sensor network data streams. Applying static restoration methods independently at each time step ignores temporal dependencies. 
As a result, the restoration quality may degrade. 
This motivates the development of restoration methods that jointly exploit spatial graph structures and temporal dynamics.

Existing methods for time-varying graph signal restoration can be broadly categorized into the following three categories. 

\textbf{Model-based methods} are generally formulated as inverse problems and solved by minimizing a handcrafted objective function consisting of data-fidelity and regularization terms that encode prior knowledge. 
In graph signal restoration, representative priors include spatial smoothness measured by the graph Laplacian quadratic form and frequency-domain constraints implemented by graph filters (e.g., low-pass/high-pass filtering), as well as temporal smoothness \cite{qiu2017TimeVarying, giraldo2022Reconstruction,isufi2019Forecasting,alikasifoglu2024Wiener,xiao2023Joint,mao2018SpatioTemporal,liu2024Adaptive}.

These formulations yield transparent interpretations of the restored signals and enable principled control of the trade-off between fidelity and smoothness via a small number of hyperparameters (e.g., regularization weights). 
On the other hand, their performance can be sensitive to the choice of the prior and hyperparameters, and hyperparameter tuning can be costly when the graph and signal statistics vary across datasets.

\textbf{Data-driven methods} learn a mapping from noisy, incomplete graph signals to the restored ones directly from training data. 
Among these, graph neural networks (GNNs), which extend convolutional neural networks to graph-structured data, have emerged as a powerful tool for graph signal restoration \cite{wu2021Comprehensive, asif2021Graph}.
Various GNN-based graph signal restoration methods have been proposed, including graph convolution using Chebyshev/Gegenbauer approximated filters \cite{he2022Convolutional, castro-correa2024Gegenbauera}, and recurrent neural networks (RNNs) combined with graph frequency analysis \cite{lewenfus2020Joint}.

These methods have been reported to show good recovery performance when the model is well-trained with a large amount of training data. 
However, collecting sufficient labeled data is often difficult in the graph setting, and the learned models may not generalize well across different graph topologies. 
Moreover, the complexity of GNN architectures can hinder interpretability, making it non-trivial to relate the learned parameters to the restoration behavior.

\textbf{Hybrid methods} have emerged as a promising direction to leverage the respective strengths of model-based and data-driven approaches. 
They integrate data-driven components into model-based frameworks, and vice versa, aiming to achieve both the interpretability of the model-based optimization and the high performance of deep learning. 

A popular class of hybrid methods includes deep algorithm unrolling (DAU) \cite{Graph_Signal_Restoration:Nagahama:TSP:2022, chen2021TimeVarying,yan2026Trainable} and plug-and-play approaches \cite{yazaki2019Interpolation}. These techniques replace certain steps or parameters in a classical iterative algorithm with learnable modules, such as neural networks.
Specifically, DAU uses deep learning techniques to determine the parameters of iterative optimization algorithms.
Each computational step of the iterative algorithm is \textit{unrolled} on the network and the parameters used in each iteration are trained using training data. 
Although convergence to the optimal solution is generally not guaranteed for iterative steps using learned parameters, practical improvements have been reported in terms of convergence speed and accuracy compared to conventional methods. 

\begin{figure}[tb]
    \centering
    \includegraphics[width=\linewidth]{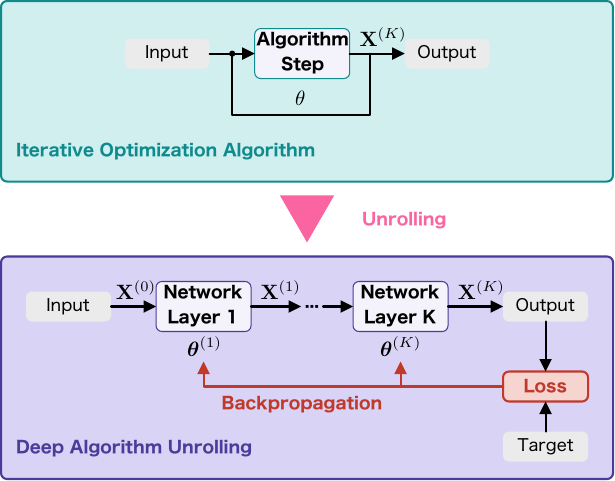}
    \caption{Comparison of conventional iterative algorithms (top) and DAU (bottom). DAU unrolls an iterative algorithm into a deep neural network, where parameters within each iteration (layer) are learned from data, leading to faster convergence and improved accuracy.}
    \label{fig:intro:dau}
\end{figure}

Figure \ref{fig:intro:dau} compares conventional iterative algorithms and DAU.
In conventional iterative optimization (shown on the top), parameters are fixed in advance and iterative steps are executed according to those parameters.
In contrast, DAU (shown on the bottom) considers each iteration as a layer of a trainable neural network and learns the parameters.
The parameters of each layer are learned from training data based on loss functions such as MSE. 
Compared with purely data-driven architectures such as GNNs, DAU-based models typically require fewer trainable parameters and preserve a clear connection to the underlying optimization problem, which improves interpretability.

Despite these advances, several challenges remain in restoring time-varying graph signals. Model-based methods rely on handcrafted priors and require careful tuning of hyperparameters, which may not generalize well across different datasets. 
Data-driven methods such as GNNs can achieve strong performance but often require large datasets and provide limited interpretability. 
Moreover, existing approaches for time-varying graph signal restoration typically impose simple and fixed temporal priors, such as first-order temporal differences. 
These priors may not capture long-term temporal dependencies that arise in real-world
signals. 
Learning such temporal structures directly from data while preserving interpretability remains an open challenge.

To address these issues, in this paper, we propose a deep algorithm unrolling (DAU) framework for time-varying graph signal restoration, termed \textbf{GLASS}\footnote{Restoration of time-varying \textbf{G}raph signals with \textbf{L}earned \textbf{A}daptive \textbf{S}patiotemporal \textbf{S}moothness.}.
The key idea is to formulate an optimization problem with a regularizer that imposes spatiotemporal smoothness.
Spatial smoothness is measured by a Sobolev operator, a higher-order graph filter that generalizes the graph Laplacian, and temporal smoothness by a multi-tap finite impulse response (FIR) filter.
We then unroll this algorithm and train both the regularization parameters and the FIR filter coefficients directly from the data. 
This allows our model to adaptively learn both the spatial and temporal characteristics of the signals, leading to more accurate restoration.
Furthermore, we extend this framework to three practical scenarios to address various real-world constraints (summarized in Table~\ref{table:intro:compare-methods}):
\begin{itemize}
    \item \textbf{Supervised Batch Method} processes all $T$ time steps of the signal sequence at once, allowing it to learn the FIR filter coefficients and regularization parameters by leveraging all available information.
    \item \textbf{Supervised Online Method} restores the signal at each time step from the current observation and the $L$ previously restored signals, where $L$ is the number of learnable FIR filter taps. It is tailored for real-time or streaming applications where low latency is critical.
    \item \textbf{Unsupervised Batch Method} processes all $T$ time steps, learning the FIR filter coefficients and the regularization parameters from the target observation alone.
    This method is developed for scenarios where no ground truth signals are available for training.
\end{itemize}

\begin{table*}[tb]
    \centering
    \caption{Comparison of existing and proposed methods.}
    \label{table:intro:compare-methods}
    \begin{tabular}{c|c||l|c|c}
    \hline
    Learning paradigm & Processing mode & Method & Category & \# of required time steps \\ \hline\hline
    \multirow{10}{*}{Supervised}
        & \multirow{4}{*}{Batch}
        & OGTR (batch) \cite{qiu2017TimeVarying}           & Model-based  & $T$ \\
        & & TRSS (batch) \cite{giraldo2022Reconstruction}  & Model-based  & $T$ \\
        & & Tikhonov \cite{perraudin2017Stationary}        & Model-based  & $T$ \\
        & & \textbf{GLASS (batch)}                          & DAU          & $T$ \\ \cline{2-5}
        & \multirow{3}{*}{Static}
        & GCN \cite{kipf2017SemiSupervised}                          & Data-driven  & $1$ \\
        & & ChebNet \cite{defferrard2016Convolutional}     & Data-driven  & $1$ \\
        & & LR \cite{smola2003Kernels}          & Model-based  & $1$ \\ \cline{2-5}
        & \multirow{2}{*}{Online}
        & OGTR (online) \cite{qiu2017TimeVarying}          & Model-based  & $2$ \\
        & & \textbf{GLASS (online)}                         & DAU          & $L+1$ \\ \hline
    \multirow{2}{*}{Unsupervised}
        & \multirow{2}{*}{Batch}
        & Inpainting DAU \cite{chen2021TimeVarying}         & DAU          & $T$ \\
        & & \textbf{GLASS (unsupervised)}                   & DAU & $T$ \\ \hline
    \end{tabular}
\end{table*}

We conduct extensive experiments on synthetic and real-world datasets to validate the effectiveness of our proposed methods.
The supervised batch method achieves the lowest RMSEs in almost all cases.
The supervised online and the unsupervised batch methods outperform the existing methods of their own settings.
Both are often competitive with the existing batch methods.

This paper is organized as follows.
Section \ref{sec:gsp} introduces the notation and the basics of graph signal processing.
Section \ref{sec:related} reviews the existing restoration methods for static and time-varying graph signals.
Section \ref{sec:proposed} presents the proposed methods, i.e., the baseline optimization problem, its unrolled network, and the three restoration settings.
Section \ref{sec:exp} shows the experimental results on several synthetic and real-world datasets.
Section \ref{sec:conclusion} concludes this paper.

\textit{Notation}: In this paper, vectors are denoted by lowercase boldface letters such as $\mathbf{x}$, and matrices are denoted by uppercase boldface letters such as $\mathbf{X}$. 
The $(i,j)$-element of the matrix $\mathbf{X}$ is denoted as $X_{i,j}$, and $\mathbf{x}_i$ denotes the $i$-th column of $\mathbf{X}$.
$\mathbf{X}_{i:j} = [\mathbf{x}_i, \mathbf{x}_{i+1}, \dots, \mathbf{x}_j]$ denotes the submatrix of the consecutive columns from the $i$-th to the $j$-th.
$\operatorname{diag}(x_0,\ldots,x_{N-1})$ denotes a diagonal matrix whose $(i,i)$-element is $x_i$.
$\mathbf{I}$ denotes the identity matrix.
$\mathcal{N}(\mu, \sigma^2)$ is a normal distribution with mean $\mu$ and variance $\sigma^2$.
Hadamard product is denoted by $\circ$.
Also, the L2 norm and Frobenius norm are denoted by $\|\cdot\|_2$ and $\|\cdot\|_F$ respectively.
$\langle \mathbf{A},\mathbf{B}\rangle = \sum_{i,j} A_{i,j}B_{i,j}$ denotes the Frobenius inner product.

\section{Basis of Graph Signal Processing} \label{sec:gsp}
In this section, we review the fundamentals of graph signal processing, including time-varying graph signals.

\subsection{Time-varying Graph Signals}
We consider a weighted undirected graph $\mathcal{G} = (\mathcal{V},\mathcal{E},\mathbf{W})$, where $\mathcal{V} = \{v_0,v_1,\ldots ,v_{N-1}\}$ is a set of $N = |\mathcal{V}|$ nodes, $\mathcal{E}$ is a set of edges, and $\mathbf{W} \in \mathbb{R}_{\ge 0}^{N\times N}$ is a weighted adjacency matrix.
The $(m,n)$-element of $\mathbf{W}$ represents the edge weight between nodes $v_m$ and $v_n$:
\begin{equation}
    W_{m,n} =
    \begin{cases}
        w_{m,n} > 0 & \text{if $v_m$ and $v_n$ are connected}, \\
        0 & \text{otherwise}.
    \end{cases}
    \label{eq:gsp:adjacency}
\end{equation}
Using $\mathbf{W}$, the combinatorial graph Laplacian of $\mathcal{G}$ is defined as
\begin{equation}
    \mathbf{L} = \mathbf{\Delta} - \mathbf{W},
    \label{eq:gsp:laplacian}
\end{equation}
where $\mathbf{\Delta}$ is the degree matrix, i.e., a diagonal matrix whose diagonal elements are given by $\Delta_{n,n} = \sum_m W_{m,n}$.
Since $\mathbf{L}$ is a real symmetric matrix, it can be diagonalized by an orthogonal matrix and thus admits the eigendecomposition
\begin{equation}
    \mathbf{L} = \mathbf{U}\mathbf{\Lambda}\mathbf{U}^\top,
    \label{eq:gsp:eig}
\end{equation}
where $\mathbf{\Lambda} = \operatorname{diag}(\lambda_0,\ldots,\lambda_{N-1})$ is the eigenvalue matrix whose diagonal entries are arranged in ascending order, and $\mathbf{U} = [\mathbf{u}_0,\ldots,\mathbf{u}_{N-1}]$ is the orthonormal matrix composed of the corresponding eigenvectors.

A graph signal is a discrete signal defined on the nodes of a graph.
For a graph with $N$ nodes, a graph signal $\mathbf{x} \in \mathbb{R}^N$ is represented as
\begin{equation}
    \mathbf{x} = [x_0,\ldots,x_{N-1}]^\top,
\end{equation}
where $x_i$ is the value observed at node $v_i$.
The smoothness of $\mathbf{x}$ on the graph is quantified by the Laplacian quadratic form
\begin{align}
    S_2(\mathbf{x}) &\coloneqq \mathbf{x}^\top\mathbf{L}\mathbf{x} \\
    &= \sum_{(m,n)\in\mathcal{E}} w_{m,n}(x_m - x_n)^2,
    \label{eq:gsp:s2}
\end{align}
which accumulates the weighted squared differences between the signal values on connected nodes.
The smaller $S_2(\mathbf{x})$ is, the more slowly the signal varies over the graph, i.e., the smoother the signal is.

When a graph signal varies over time, it is called a time-varying graph signal.
By stacking the graph signals at $T$ consecutive time instances as columns, a time-varying graph signal is expressed as
\begin{equation}
    \mathbf{X} = [\mathbf{x}_0,\dots, \mathbf{x}_{T-1}] \in \mathbb{R}^{N \times T},
    \label{eq:gsp:tvgs}
\end{equation}
where $T$ is the number of time instances and $\mathbf{x}_t$ is the graph signal observed at time $t$.

\section{Related Work} \label{sec:related}
In this section, we introduce existing static and time-varying graph signal restoration methods. 

\subsection{Restoration of Static Graph Signals}
Graph signals are often corrupted by noise during measurement and transmission.
Some nodes are also left unobserved when sensors or communication links fail.
The observed static graph signal $\mathbf{y} \in \mathbb{R}^{N}$ can be modeled as follows:
\begin{equation}
\mathbf{y} = \mathbf{j} \circ (\mathbf{x}^* + \mathbf{v}),\label{eq:existing:static-modeling}
\end{equation}
where $\mathbf{x}^* \in \mathbb{R}^{N}$ is the (unknown) original graph signal, $\mathbf{v} \in \mathbb{R}^{N}$ is the noise term, and $\mathbf{j} \in \mathbb{R}^{N}$ is the measurement vector that satisfies
\begin{equation}
\begin{split}
[\mathbf{j}]_n =
\begin{cases}
1 & \text{if a signal value is observed on node } n,\\
0 & \text{otherwise}.
\end{cases}\label{eq:existing:samplingop}
\end{split}
\end{equation}
Here, the measurement vector $\mathbf{j}$ is assumed to be known a priori, i.e., the positions of the observed entries are given.

For static graph signal restoration, Tikhonov regularization \cite{shuman2013emerging} is widely used.
It assumes the spatial smoothness of the graph signal and restores it by solving the following minimization problem:
\begin{equation}
\Tilde{\mathbf{x}} = \argmin _{\mathbf{x}} \frac{1}{2}\|\mathbf{j}\circ \mathbf{x}-\mathbf{y}\|^2_2+ \frac{\alpha}{2}\|\mathbf{H}\mathbf{x}\|^2_2, \label{eq:existing:tikhonov}
\end{equation}
where $\alpha$ is a regularization parameter and $\mathbf{H} \in \mathbb{R}^{M\times N}$ is a regularization operator.
The first term of \eqref{eq:existing:tikhonov} represents the data fidelity between the observed signal and the restored signal.
In graph signal processing, $\mathbf{H}$ is usually chosen as a graph high-pass filter, so that the second term measures the high-frequency components of the restored graph signal.
Minimizing this objective suppresses the high-frequency components, i.e., it makes the restored graph signal smooth.

Laplacian regularization \cite{smola2003Kernels} is a special case of \eqref{eq:existing:tikhonov} with $\mathbf{H}=\mathbf{L}^{1/2}$.
This substitution gives $\|\mathbf{H}\mathbf{x}\|^2_2 = \mathbf{x}^\top\mathbf{L}\mathbf{x} = S_2(\mathbf{x})$ in \eqref{eq:gsp:s2}, and the minimization problem is given as
\begin{equation}
\Tilde{\mathbf{x}} = \argmin _{\mathbf{x}} \frac{1}{2}\|\mathbf{j}\circ \mathbf{x}-\mathbf{y}\|^2_2+ \frac{\alpha}{2} S_2(\mathbf{x}). \label{eq:existing:laplacian}
\end{equation}
For a connected graph, $S_2(\mathbf{x})=0$ holds only when $\mathbf{x}=c\mathbf{1}$ with a constant $c$.
This regularization thus restores the graph signal to be spatially smooth over the graph.

Since these methods are for static graph signals, applying them to the time-varying settings requires solving the optimization problem independently at each time $t=0,\dots,T-1$.
Consequently, they cannot exploit the temporal relationships between consecutive signals, which may degrade the restoration performance.

\subsection{Restoration of Time-varying Graph Signals}
Similar to the static case in \eqref{eq:existing:static-modeling}, an observed time-varying graph signal $\mathbf{Y} = [\mathbf{y}_0,\dots,\mathbf{y}_{T-1}] \in \mathbb{R}^{N\times T}$ can be modeled as follows:
\begin{equation}
\mathbf{Y} = \mathbf{J} \circ (\mathbf{X}^* + \mathbf{V}), \label{eq:existing:timevarying-modeling}
\end{equation}
where $\mathbf{X}^* \in \mathbb{R}^{N \times T}$ is the (unknown) original time-varying graph signal, $\mathbf{V} \in \mathbb{R}^{N \times T}$ is the noise term, and $\mathbf{J} \in \mathbb{R}^{N \times T}$ is the measurement matrix that satisfies the following condition, similar to \eqref{eq:existing:samplingop}:
\begin{equation}
    \begin{split}
    J_{n,t} =
        \begin{cases}
        1 & \text{if a signal value is observed on node } n \text{ at time } t,\\
        0 & \text{otherwise}.\label{eq:existing:sampling}
        \end{cases}
    \end{split}
\end{equation}
As in the static case, the measurement matrix $\mathbf{J}$ is assumed to be known a priori.

Spatiotemporal regularization \cite{qiu2017TimeVarying} is a representative method for restoring time-varying graph signals.
In \cite{qiu2017TimeVarying}, the Laplacian quadratic form for a time-varying graph signal is defined as
\begin{align}
    S_2(\mathbf{X}) &\coloneqq \sum_{t=0}^{T-1} \mathbf{x}_t^\top\mathbf{L}\mathbf{x}_t\\
                    &= \operatorname{tr}(\mathbf{X}^\top\mathbf{L}\mathbf{X}). \label{eq:existing:s2timevarying}
\end{align}
Using \eqref{eq:existing:s2timevarying}, the optimization problem in \cite{qiu2017TimeVarying} is modeled as
\begin{equation}
    \Tilde{\mathbf{X}} = \argmin _{\mathbf{X}} \frac{1}{2}\|\mathbf{J}\circ \mathbf{X}-\mathbf{Y}\|^2_F+\frac{\alpha}{2} \operatorname{tr}\left(\left(\mathbf{X}\mathbf{D}\right)^{\top} \mathbf{L} \left(\mathbf{X} \mathbf{D}\right)\right), \label{eq:existing:OGTR}
\end{equation}
where $\alpha$ is a hyperparameter that controls the smoothness of the restored signal, and $\mathbf{D} \in \mathbb{R}^{T\times (T-1)}$ is the temporal two-tap FIR filter defined as
\begin{equation}
\mathbf{D}=\left[\begin{array}{cccc}-1 & & &\\
1 & -1 & &\\
& 1 & \ddots &\\
& & \ddots & -1\\
& & & 1 \end{array}\right] \in \mathbb{R}^{T\times (T-1)}.
\end{equation}
In \eqref{eq:existing:OGTR}, the first term enforces data fidelity, while the second term promotes spatiotemporal smoothness.
Here, $\mathbf{X}\mathbf{D} = [\mathbf{x}_1 - \mathbf{x}_0, \mathbf{x}_2 - \mathbf{x}_1, \dots, \mathbf{x}_{T-1} - \mathbf{x}_{T-2}]$ is the temporal difference between graph signals at consecutive time instances.
Applying the Laplacian quadratic form to it penalizes signals that are non-smooth in both the spatial and temporal domains.
However, \eqref{eq:existing:OGTR} relies on low-order operators in both domains: the graph Laplacian $\mathbf{L}$ only accounts for one-hop spatial relationships, and the fixed two-tap filter $\mathbf{D}$ only accounts for the difference between two consecutive time instances.

To address the spatial limitation, \cite{giraldo2022Reconstruction} replaces the graph Laplacian $\mathbf{L}$ in \eqref{eq:existing:OGTR} with the Sobolev operator $(\mathbf{L}+\epsilon\mathbf{I})^{\beta}$, and models the optimization problem as
\begin{equation}
    \begin{split}
    \Tilde{\mathbf{X}} = \argmin _{\mathbf{X}} \; & \frac{1}{2}\|\mathbf{J}\circ \mathbf{X}-\mathbf{Y}\|^2_F \\
    & + \frac{\alpha}{2} \operatorname{tr}\left(\left(\mathbf{X}\mathbf{D}\right)^{\top} (\mathbf{L}+\epsilon\mathbf{I})^{\beta} \left(\mathbf{X} \mathbf{D}\right)\right),
    \end{split}
    \label{eq:existing:TRSS}
\end{equation}
where $\alpha \geq 0$ and $\mathbf{D}$ are the same as in \eqref{eq:existing:OGTR}, and $\epsilon > 0$ and $\beta \geq 0$ are hyperparameters. Since $\mathbf{L}$ and $(\mathbf{L}+\epsilon\mathbf{I})^{\beta}$ share their eigenvectors, the Sobolev operator only reshapes the spectral response from $\lambda$ to $(\lambda+\epsilon)^{\beta}$. 
$\epsilon$ makes the operator positive definite, so the null space of $\mathbf{L}$ is also penalized. $\beta$ changes the shape of the response.
At $\beta=1$, it is the response of $\mathbf{L}$ shifted by $\epsilon$.
A larger $\beta$ penalizes the high graph frequencies more strongly.
For $\beta=0$, all the graph frequencies are penalized equally.
The spectral response of the regularizer can thus be adjusted by $\epsilon$ and $\beta$.
The temporal filter $\mathbf{D}$, in contrast, is fixed to the two-tap difference.
The temporal smoothness is therefore captured only between consecutive time instances.

\section{GLASS} \label{sec:proposed}

\begin{figure*}[tb]
    \centering
    \includegraphics[width=\textwidth]{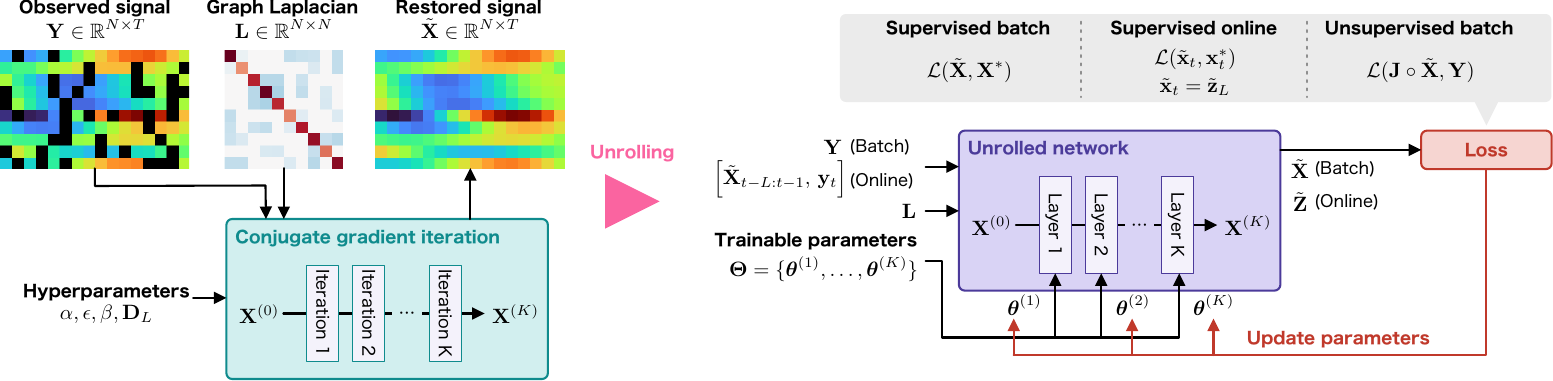}
    \caption{Overview of the baseline method (left) and the proposed GLASS (right). The three settings of GLASS share the unrolled network and differ only in the input, the output, and the loss. In the online setting, the network takes the window $[\Tilde{\mathbf{X}}_{t-L:t-1}, \mathbf{y}_t]$ and outputs $\Tilde{\mathbf{Z}}$. Its last column $\Tilde{\mathbf{z}}_L$ is the restored frame $\Tilde{\mathbf{x}}_t$.}
    \label{fig:proposed:overview}
\end{figure*}

In this section, we introduce our proposed method for restoring time-varying graph signals using DAU. 
First, we formulate an optimization problem using an FIR filter for temporal smoothness and the Sobolev operator for spatial smoothness.
Next, an iterative algorithm for solving the formulated optimization problem is \textit{unrolled} with DAU. 
Then, we train the parameters of the unrolled network in a supervised or unsupervised manner, yielding three restoration settings: supervised batch, supervised online, and unsupervised batch.
Finally, we verify the applicability of DAU to the proposed baseline problem by validating the differentiability of the algorithm and loss function.

\subsection{Baseline Problem}
First, we consider the following optimization problem as an extension of \eqref{eq:existing:TRSS}:
\begin{equation}
    \begin{split}
    \Tilde{\mathbf{X}} = \argmin _{\mathbf{X}} \; & \frac{1}{2}\|\mathbf{J}\circ \mathbf{X}-\mathbf{Y}\|^2_F \\
    & + \frac{\alpha}{2} \operatorname{tr}\left(\left(\mathbf{X}\mathbf{D}_{L}\right)^{\top} (\mathbf{L}+\epsilon\mathbf{I})^{\beta} \left(\mathbf{X} \mathbf{D}_{L}\right)\right)
    \end{split}
    \label{eq:proposed:baseline}
\end{equation}
where $\alpha \geq 0$, $\epsilon > 0$, and $\beta \geq 0$ are hyperparameters; $\alpha$ determines the smoothness of the time-varying graph signals, and $(\mathbf{L}+\epsilon\mathbf{I})^{\beta}$ is the Sobolev operator. $\mathbf{D}_L$ is a causal FIR filter with $L+1$ taps defined as
\begin{equation}
    \label{eq:proposed:d_L_operator}
    \begin{split}
        \mathbf{D}_{L} = 
        \begin{bmatrix}
            -d_L    &        &      &\\
            -d_{L-1}&-d_L    &      &\\
            \vdots  &\vdots  &      &\\
            -d_1    &-d_2    &      &\\
            1       &-d_1    &\ddots&\vdots\\
                    &1       &\ddots&-d_2\\
                    &        &\ddots&-d_1\\
                    &        &      &1\\
        \end{bmatrix} \in \mathbb{R}^{T \times (T-L)}.
    \end{split}
\end{equation}
The coefficients $\mathbf{d} = [d_1,\ldots,d_L]\in \mathbb{R}_{\ge 0}^L$ satisfy $\sum_{l=1}^L d_l = 1$.
The first term in \eqref{eq:proposed:baseline} is the well-known data-fidelity term.
The second term applies the Sobolev operator to the filtered signal $\mathbf{X}\mathbf{D}_L$.
It therefore promotes small spatiotemporal variations.
Compared with \eqref{eq:existing:TRSS}, the fixed two-tap filter $\mathbf{D}$ is replaced by the $(L+1)$-tap filter $\mathbf{D}_L$.
The temporal smoothness is thus measured over multiple past frames.
If $L=1$, \eqref{eq:proposed:baseline} coincides with \eqref{eq:existing:TRSS}.

A solution of \eqref{eq:proposed:baseline} can be obtained by an iterative solver such as the conjugate gradient (CG) method, as summarized in Algorithm \ref{alg:proposed:baseline}.
In Algorithm \ref{alg:proposed:baseline}, $\Delta \mathbf{X}^{(k)}$ and $\nabla\mathbf{X}^{(k)}$ are calculated as follows:
\begin{align}
    \Delta \mathbf{X}^{(k)} &= - \nabla\mathbf{X}^{(k)}+ \gamma \Delta \mathbf{X}^{(k-1)}\\
    \nabla\mathbf{X}^{(k)} &= \mathbf{J} \circ \mathbf{X}^{(k)} - \mathbf{Y} + \alpha (\mathbf{L}+\epsilon\mathbf{I})^{\beta}\mathbf{X}^{(k)}\mathbf{D}_{L}\mathbf{D}_{L}^{\top}. \label{eq:proposed:gradient}
\end{align}
An overview of the baseline method is shown on the left of Fig. \ref{fig:proposed:overview}.
We can thus obtain $\Tilde{\mathbf{X}}$ using Algorithm \ref{alg:proposed:baseline} if the parameters $\alpha$, $\epsilon$, $\beta$, and $\mathbf{D}_L$ are fixed in advance.

\begin{algorithm}[t]
    \caption{Conjugate Gradient Method for the Baseline Problem~\eqref{eq:proposed:baseline}}
    \label{alg:proposed:baseline}
    \begin{algorithmic}

    \Require{\\
      $\mathbf{Y}$: Observed signal\\
      $\mathbf{J}$: Measurement matrix defined in \eqref{eq:existing:sampling}\\
      $\mathbf{L}$: Graph Laplacian\\
      $\alpha,\epsilon,\beta$: Regularization parameters\\
      $\mathbf{D}_L$: FIR filter\\
      $K$: Number of iterations\\
      $\delta$: Small positive constant}
    \Ensure{\\$\tilde{\mathbf{X}} = \mathbf{X}^{(K)}$: Restored signal}\\
    \hrulefill
    \State{1) Initialization}
    \State{~~~ $\mathbf{X}^{(0)} = \mathbf{0}$}
    \State{~~~ $\Delta \mathbf{X}^{(0)} = - \nabla\mathbf{X}^{(0)}$}
    \For{$k = 1$ to $K$}
        \State{2) Stepsize decision}
        \State{~~~ $\tau = - \frac{\langle \Delta \mathbf{X}^{(k-1)} , \nabla \mathbf{X}^{(k-1)} \rangle }{\langle \Delta \mathbf{X}^{(k-1)} , \mathbf{J}\circ\Delta \mathbf{X}^{(k-1)} + \alpha (\mathbf{L}+\epsilon\mathbf{I})^{\beta}\Delta \mathbf{X}^{(k-1)}\mathbf{D}_{L}\mathbf{D}_{L}^{\top} \rangle + \delta}$}
        \State{3) Search direction update}
        \State{~~~ $\mathbf{X}^{(k)} = \mathbf{X}^{(k-1)} + \tau \Delta \mathbf{X}^{(k-1)}$}
        \State{~~~ $\gamma = \frac{\left\lVert \nabla\mathbf{X}^{(k)} \right\rVert_F^2 }{\left\lVert \nabla\mathbf{X}^{(k-1)} \right\rVert_F^2 + \delta}$}
        \State{~~~ $\Delta \mathbf{X}^{(k)} = - \nabla\mathbf{X}^{(k)} + \gamma \Delta \mathbf{X}^{(k-1)}$}
    \EndFor
    \end{algorithmic}
\end{algorithm}

\subsection{Proposed Restoration with DAU} \label{sec:proposed-dau}

The restoration quality of Algorithm \ref{alg:proposed:baseline} depends on the parameters $\alpha$, $\epsilon$, $\beta$, and $\mathbf{D}_L$.
They should be fitted according to the characteristics of the data, the noise level, and the missing ratio.
Their appropriate values, however, are difficult to determine in advance.

An overview of the proposed method is shown on the right of Fig. \ref{fig:proposed:overview}.
In the proposed method, the iterative steps of the conjugate gradient method are \textit{unrolled} on a network by DAU, and the network is trained by supervised / unsupervised learning to obtain optimal parameters.
The proposed DAU network consists of all $K$ layers (i.e., $K$ iterations). 
The network takes the observed signal $\mathbf{Y}$, the measurement matrix $\mathbf{J}$, and the graph Laplacian $\mathbf{L}$.
After $K$ iterations, it outputs the restored signal $\Tilde{\mathbf{X}} = \mathbf{X}^{(K)}$.
The $k$-th layer takes the restored signal from the previous layer $\mathbf{X}^{(k-1)}$ and performs one CG iteration of Algorithm \ref{alg:proposed:baseline}.
In this iteration, the parameters $\alpha$, $\epsilon$, $\beta$, and $\mathbf{D}_L$ are replaced with a layer-specific trainable parameter set $\boldsymbol{\theta}^{(k)} = \{\boldsymbol{\rho}^{(k)}, \mathbf{d}^{(k)}\}$.
Here, $\boldsymbol{\rho}^{(k)} = \{\alpha^{(k)},\epsilon^{(k)},\beta^{(k)}\}$ is a collection of the regularization parameters, and $\mathbf{d}^{(k)} = [d_1^{(k)},\ldots,d_L^{(k)}]$ represents the FIR filter coefficients.
Collecting the parameters of all layers as $\boldsymbol{\Theta} = \{\boldsymbol{\theta}^{(1)},\dots,\boldsymbol{\theta}^{(K)}\}$, we denote the forward pass of the unrolled network by $\mathrm{GLASS}(\mathbf{Y},\mathbf{J},\mathbf{L},\boldsymbol{\Theta})$, which outputs the restored signal $\Tilde{\mathbf{X}} = \mathbf{X}^{(K)}$.

\subsection{Supervised Restoration} \label{sec:proposed-supervised}
Here, we show the supervised batch/online restoration method for learning parameters using a training dataset.
The supervised batch method reconstructs from the observed signal all at once.
The supervised online method uses the current observation and a few previously restored frames.

\subsubsection{Supervised Batch Method}
The supervised batch method uses the whole observed signal $\mathbf{Y}$ to compute the restored signal $\tilde{\mathbf{X}}$.
The parameters $\boldsymbol{\Theta}$ are updated using the training dataset.
We use the mean squared error (MSE) as the loss function, defined as
\begin{equation}
    \mathcal{L}(\mathbf{X}^*,\Tilde{\mathbf{X}}) = \frac{1}{NT} \|\Tilde{\mathbf{X}}-\mathbf{X}^*\|^2_F, \label{eq:proposed-supervised-mse}
\end{equation}
where $\mathbf{X}^*$ is the ground truth and $\Tilde{\mathbf{X}}$ is the output of the restoration model.
By utilizing the parameters learned from the training data, it is possible to achieve high-quality restoration for the test data.

\subsubsection{Supervised Online Method}
The supervised batch method is expected to have high restoration performance, but it also introduces delay.
To avoid this delay in real-time or streaming applications, we propose an online restoration method that restores each frame in a causal manner.
It uses only the current observation $\mathbf{y}_t$ and the $L$ previously restored frames $\Tilde{\mathbf{X}}_{t-L:t-1} \in \mathbb{R}^{N\times L}$.
At time $t$, we form the window $[\Tilde{\mathbf{X}}_{t-L:t-1}, \mathbf{y}_t] \in \mathbb{R}^{N\times (L+1)}$ and apply the unrolled network once, taking its last column as the restored frame:
\begin{align}
    \Tilde{\mathbf{x}}_t &= \Tilde{\mathbf{z}}_L, \\
    \Tilde{\mathbf{Z}} &= \mathrm{GLASS}([\Tilde{\mathbf{X}}_{t-L:t-1}, \mathbf{y}_t],[\mathbf{1},\dots,\mathbf{1},\mathbf{j}_t],\mathbf{L},\boldsymbol{\Theta}), \label{eq:proposed-online}
\end{align}
where the $L$ past restored frames are input, and only the current frame keeps its measurement vector $\mathbf{j}_t$.
For $t<L$, some of the $L$ past frames do not exist yet.
Their columns are set to $\mathbf{0}$ in both the window and the measurement matrix, i.e., they are treated as unobserved.
The parameters $\boldsymbol{\Theta}$ are trained for this recursion by supervised learning with the loss function in \eqref{eq:proposed-supervised-mse}.

\subsection{Unsupervised Batch Restoration} \label{sec:proposed-unsupervised}
The supervised method introduced in Section \ref{sec:proposed-supervised} requires (a small number of) datasets for training. 
However, there may be cases where no ground truth is available for the target data, such as restoring data with unknown features.
For such situations, we propose an unsupervised batch restoration method that requires no ground truth of the target data.

In the unsupervised restoration, MSE loss is calculated as follows:
\begin{equation}
    \mathcal{L}(\mathbf{Y},\Tilde{\mathbf{X}}) = \frac{1}{NT} \|\mathbf{J} \circ (\Tilde{\mathbf{X}}-\mathbf{Y})\|^2_F. \label{eq:proposed-unsupervised-mse}
\end{equation}
Since the ground truth signal is unavailable, \eqref{eq:proposed-unsupervised-mse} compares the restored signal with the observation $\mathbf{Y}$ only at the observed entries through the Hadamard product $\mathbf{J}\circ\Tilde{\mathbf{X}}$.
Hence, training requires only the observed signal $\mathbf{Y}$ of the target data and no ground truth.

\subsection{Differentiability of the Algorithm}
Our proposed method trains the parameters $\boldsymbol{\Theta}$ by gradient descent through DAU.
This is possible only if the iterative update steps of Algorithm \ref{alg:proposed:baseline} and the loss function $\mathcal{L}$ are differentiable with respect to $\boldsymbol{\Theta}$.
Here, we verify this differentiability.

Algorithm \ref{alg:proposed:baseline} consists only of sums, Hadamard products, matrix products, inner products, and two quotients.
All of these operations are differentiable, as long as the denominator of a quotient is nonzero.
It is therefore enough to find the parts that depend on $\boldsymbol{\Theta}$, and to check them one by one.

In the $k$-th layer, the parameters $\boldsymbol{\theta}^{(k)}$ appear in two places: the gradient \eqref{eq:proposed:gradient} and the denominator of the stepsize $\tau$.
Both contain the parameters only through $\alpha^{(k)}$, the Sobolev operator $(\mathbf{L}+\epsilon^{(k)}\mathbf{I})^{\beta^{(k)}}$, and $\mathbf{D}_{L}^{(k)}\mathbf{D}_{L}^{(k)\top}$.
These three are the parts to check in each layer.

Both \eqref{eq:proposed:gradient} and the denominator of $\tau$ are linear in $\alpha^{(k)}$.
Each entry of the FIR filter \eqref{eq:proposed:d_L_operator} is $1$, $0$, or $-d_l^{(k)}$.
Hence each entry of $\mathbf{D}_{L}^{(k)}\mathbf{D}_{L}^{(k)\top}$ is a quadratic polynomial in $d_1^{(k)},\ldots,d_L^{(k)}$.
These two parts are thus polynomial in the parameters, and their derivatives are immediate.

The remaining part is the Sobolev operator.
Using the eigendecomposition \eqref{eq:gsp:eig}, we write it as
\begin{equation}
    (\mathbf{L}+\epsilon^{(k)}\mathbf{I})^{\beta^{(k)}} = \mathbf{U}(\mathbf{\Lambda}+\epsilon^{(k)}\mathbf{I})^{\beta^{(k)}}\mathbf{U}^{\top}. \label{eq:proposed:sobolev-eig}
\end{equation}
In \eqref{eq:proposed:sobolev-eig}, only the eigenvalues $(\lambda_i+\epsilon^{(k)})^{\beta^{(k)}}$ depend on $\epsilon^{(k)}$ and $\beta^{(k)}$.
The derivatives of the operator are therefore
\begin{align}
\frac{\partial (\mathbf{L}+\epsilon^{(k)}\mathbf{I})^{\beta^{(k)}}}{\partial \epsilon^{(k)}}&= \beta^{(k)}(\mathbf{L}+\epsilon^{(k)}\mathbf{I})^{\beta^{(k)}-1},\label{eq:proposed:dS-e}\\
\frac{\partial (\mathbf{L}+\epsilon^{(k)}\mathbf{I})^{\beta^{(k)}}}{\partial \beta^{(k)}}&= \mathbf{U}\operatorname{diag}\!\bigl(\eta_i \ln(\lambda_i+\epsilon^{(k)})\bigr)\mathbf{U}^{\top},\label{eq:proposed:dS-be}
\end{align}
where $\eta_i := (\lambda_i+\epsilon^{(k)})^{\beta^{(k)}}$.
Here, $\lambda_i+\epsilon^{(k)}>0$ holds because $\epsilon^{(k)}>0$ and $\lambda_i\geq 0$.
The logarithm in \eqref{eq:proposed:dS-be} is thus finite, and both derivatives are well defined.

Finally, the loss functions \eqref{eq:proposed-supervised-mse} and \eqref{eq:proposed-unsupervised-mse} are quadratic in $\Tilde{\mathbf{X}}$, and hence differentiable.

Both Algorithm \ref{alg:proposed:baseline} and the loss are therefore differentiable with respect to all trainable parameters.
Our baseline problem \eqref{eq:proposed:baseline} is thus \textit{unrollable} by DAU.

\section{Experiments} \label{sec:exp}
In this section, we compare our proposed supervised batch, supervised online, and unsupervised batch time-varying graph signal restoration methods with existing methods\footnote{The Python code for the existing and proposed methods and for all experiments will be available at \url{https://github.com/kojima-msp/GLASS}}. 

\subsection{Comparison Methods}
To evaluate our method, we compare the restoration performance using several existing GSP methods.
An overview of the proposed and comparison methods is summarized in Table \ref{table:exp:compare-methods}.

For the batch setting, we use OGTR (batch) \cite{qiu2017TimeVarying} and TRSS (batch) \cite{giraldo2022Reconstruction}, which solve \eqref{eq:existing:OGTR} and \eqref{eq:existing:TRSS}, respectively.
We also use Tikhonov regularization in the joint time-vertex domain \cite{perraudin2017Stationary}, denoted by Tikhonov.

As static methods, we use Laplacian regularization (LR) \cite{smola2003Kernels} and two graph neural networks, GCN \cite{kipf2017SemiSupervised} and ChebNet \cite{defferrard2016Convolutional}.

For the online setting, we use the online variant of OGTR \cite{qiu2017TimeVarying}, denoted by OGTR (online).
For the unsupervised setting, we use Inpainting DAU \cite{chen2021TimeVarying}, which learns filters in the spatiotemporal domain.

Table \ref{table:exp:compare-methods} also summarizes the learning paradigm and the parameters of each method.
The model-based methods have a few hyperparameters, which we fit by Bayesian optimization \cite{akiba2019Optuna} on the training data.
The remaining methods are trained on the training data by Adam \cite{DBLP:journals/corr/KingmaB14} for 100 epochs in the batch setting.
In the online setting, one epoch takes much longer, so we use 50 epochs.
The two unsupervised methods use no training data.
They train their parameters on the test observation $\mathbf{Y}$ itself for 100 epochs.

\begin{table*}[tb]
\centering
    \centering
    \caption{Overview of the existing and proposed methods. $P$ and $Q$ in ``Inpainting DAU" denote the number of filter taps in the time and spatial domain, respectively.}
    \label{table:exp:compare-methods}
    \begin{tabular}{c|c||l|c|c|c}
    \hline
    Learning paradigm & types & Methods & \# of iterations / layers & parameter(s) & \# of parameters(s) \\ \hline\hline
    \multirow{9}{*}{Supervised}
    & \multirow{4}{*}{Batch} & OGTR (batch) \cite{qiu2017TimeVarying} & $K=50$ & hyperparameter & $1$ \\
    & & TRSS (batch) \cite{giraldo2022Reconstruction} & $K=50$ & hyperparameter & $3$ \\
    & & Tikhonov \cite{perraudin2017Stationary} & $K=50$ & hyperparameter & $2$ \\
    & & \textbf{GLASS (batch)} & $K=50$ & train & $K\times (L+3)$ \\ \cline{2-6}
    & \multirow{3}{*}{Static} & GCN \cite{kipf2017SemiSupervised} & $2$ & train & $129$ \\
    & & ChebNet \cite{defferrard2016Convolutional} & $2$ & train & $321$ \\
    & & LR \cite{smola2003Kernels} & $K=50$ & hyperparameter & $1$ \\ \cline{2-6}
    & \multirow{2}{*}{Online} & OGTR (online) \cite{qiu2017TimeVarying} & $K=50$ & hyperparameter & $1$ \\
    & & \textbf{GLASS (online)} & $K=50$ & train & $K\times (L+3)$ \\ \hline
    \multirow{2}{*}{Unsupervised}
    & \multirow{2}{*}{Batch} & Inpainting DAU \cite{chen2021TimeVarying} & $K=50$ & train & $K\times (P+Q-1)$ \\
    & & \textbf{GLASS (unsupervised)} & $K=50$ & train & $K\times (L+3)$ \\ \hline
    \end{tabular}
\end{table*}

\subsection{Setup}
We compare the restoration performance of time-varying graph signals for various sampling ratios.
In each experiment, we simulate the missing observations and the noise as $\mathbf{Y} = \mathbf{J}_\psi \circ (\mathbf{X}^* + \mathbf{V}_\eta)$, following \eqref{eq:existing:timevarying-modeling}.
The measurement matrix $\mathbf{J}_\psi\in\{0,1\}^{N\times T}$ is a random matrix with $\psi NT$ ones, where $\psi$ is the sampling ratio.
$\mathbf{V}_\eta$ is white Gaussian noise with entries drawn from $\mathcal{N}(0,(\eta\sigma)^2)$, where $\sigma$ is the standard deviation of $\mathbf{X}^*$ and $\eta$ is the noise level.
Each supervised method is trained on all combinations of $\psi \in \{0.1,0.2,\dots,0.9\}$ and $\eta \in \{0.02,0.04,\dots,0.10\}$.

For evaluation, we fix the noise level to $\eta=0.1$ and vary $\psi$ over the same range.
The number of learnable FIR taps of the proposed methods is set to $L=8$, which is chosen in Section \ref{sec:exp-L}.
Restoration performance is evaluated by the root mean square error (RMSE)
\begin{equation}
    \operatorname{RMSE}(\mathbf{X}^*,\tilde{\mathbf{X}}) = \sqrt{\frac{1}{NT}\|\tilde{\mathbf{X}} - \mathbf{X}^*\|_F^2},
\end{equation}
where $\mathbf{X}^*$ is the ground truth and $\tilde{\mathbf{X}}$ is a restored time-varying graph signal.

\begin{table*}[t]
    \centering
    \caption{RMSE for various sampling ratios $\psi$. B, S, O, and U denote the batch, static, online, and unsupervised methods, respectively. The best and the second best in each group are shown in bold and underlined. The best and the second best over all the methods are colored in red and blue.}
    \label{tab:ratio}
    \begin{tabular}{l||ccccccccc}
\hline
Methods $\backslash$ Metrics & $\psi=0.1$ & 0.2 & 0.3 & 0.4 & 0.5 & 0.6 & 0.7 & 0.8 & 0.9 \\ \hline\hline
\multicolumn{10}{c}{(a) Synthetic} \\ \hline
B: OGTR & 9.72&6.26&4.09&2.73&1.95&1.49&1.26&1.15&1.09 \\
B: TRSS & 2.41&1.38&1.15&\textcolor{blue}{\underline{1.04}}&\textcolor{blue}{\underline{0.96}}&\textcolor{blue}{\underline{0.90}}&\textcolor{blue}{\underline{0.86}}&\textcolor{blue}{\underline{0.82}}&\textcolor{blue}{\underline{0.79}} \\
B: Tikhonov & \underline{2.13}&\textcolor{blue}{\underline{1.36}}&\textcolor{blue}{\underline{1.14}}&\textcolor{blue}{\underline{1.04}}&\textcolor{blue}{\underline{0.96}}&\textcolor{blue}{\underline{0.90}}&\textcolor{blue}{\underline{0.86}}&\textcolor{blue}{\underline{0.82}}&\textcolor{blue}{\underline{0.79}} \\
B: \textbf{GLASS (batch)} & \textcolor{red}{\textbf{1.70}}&\textcolor{red}{\textbf{1.22}}&\textcolor{red}{\textbf{1.07}}&\textcolor{red}{\textbf{0.96}}&\textcolor{red}{\textbf{0.88}}&\textcolor{red}{\textbf{0.82}}&\textcolor{red}{\textbf{0.77}}&\textcolor{red}{\textbf{0.74}}&\textcolor{red}{\textbf{0.71}} \\ \hline
S: GCN & 12.19&10.41&8.91&7.62&6.54&5.60&4.72&3.85&2.94 \\
S: ChebNet & \textbf{11.40}&\underline{9.15}&\underline{7.62}&\underline{6.41}&\underline{5.46}&\underline{4.67}&\underline{3.94}&\underline{3.23}&\underline{2.46} \\
S: LR & \underline{11.46}&\textbf{8.88}&\textbf{7.30}&\textbf{6.12}&\textbf{5.22}&\textbf{4.47}&\textbf{3.79}&\textbf{3.11}&\textbf{2.38} \\ \hline
O: OGTR (online) & \underline{4.38}&\underline{2.28}&\underline{1.74}&\underline{1.53}&\underline{1.41}&\underline{1.34}&\underline{1.28}&\underline{1.23}&\underline{1.19} \\
O: \textbf{GLASS (online)} & \textbf{2.57}&\textbf{1.89}&\textbf{1.58}&\textbf{1.39}&\textbf{1.24}&\textbf{1.13}&\textbf{1.04}&\textbf{0.96}&\textbf{0.90} \\ \hline
U: Inpainting\_DAU & \underline{13.18}&\underline{11.27}&\underline{9.87}&\underline{8.76}&\underline{7.35}&\underline{6.44}&\underline{5.11}&\underline{4.15}&\underline{2.89} \\
U: \textbf{GLASS (unsupervised)} & \textcolor{blue}{\textbf{2.06}}&\textbf{1.57}&\textbf{1.38}&\textbf{1.25}&\textbf{1.17}&\textbf{1.11}&\textbf{1.07}&\textbf{1.02}&\textbf{0.98} \\
\hline
\multicolumn{10}{c}{(b) Japan COVID-19 (RMSE $/100$)} \\ \hline
B: OGTR & 14.23&9.63&8.37&7.39&6.67&5.95&5.29&4.61&4.09 \\
B: TRSS & \textcolor{blue}{\underline{10.83}}&\textcolor{blue}{\underline{8.75}}&\textcolor{blue}{\underline{7.85}}&\textcolor{blue}{\underline{6.93}}&\textcolor{blue}{\underline{6.30}}&\textcolor{blue}{\underline{5.56}}&\textcolor{blue}{\underline{4.87}}&\textcolor{blue}{\underline{4.17}}&\underline{3.61} \\
B: Tikhonov & 11.53&9.53&8.73&8.00&7.57&7.01&6.54&6.06&5.69 \\
B: \textbf{GLASS (batch)} & \textcolor{red}{\textbf{9.69}}&\textcolor{red}{\textbf{7.63}}&\textcolor{red}{\textbf{6.58}}&\textcolor{red}{\textbf{5.78}}&\textcolor{red}{\textbf{5.23}}&\textcolor{red}{\textbf{4.62}}&\textcolor{red}{\textbf{4.09}}&\textcolor{red}{\textbf{3.57}}&\textcolor{red}{\textbf{3.09}} \\ \hline
S: GCN & \underline{25.15}&\underline{23.32}&\underline{21.65}&19.90&18.24&16.32&14.15&11.70&\underline{8.62} \\
S: ChebNet & \textbf{23.79}&\textbf{21.33}&\textbf{19.37}&\textbf{17.54}&\textbf{15.78}&\textbf{14.11}&\textbf{12.46}&\underline{10.92}&9.36 \\
S: LR & 29.25&25.47&22.65&\underline{19.79}&\underline{17.76}&\underline{15.51}&\underline{13.16}&\textbf{10.57}&\textbf{7.62} \\ \hline
O: OGTR (online) & \underline{16.76}&\underline{10.92}&\underline{8.91}&\underline{7.77}&\underline{7.13}&\underline{6.39}&\underline{5.77}&\underline{5.16}&\underline{4.60} \\
O: \textbf{GLASS (online)} & \textbf{14.06}&\textbf{10.21}&\textbf{8.33}&\textbf{7.22}&\textbf{6.41}&\textbf{5.59}&\textbf{4.88}&\textcolor{blue}{\textbf{4.17}}&\textcolor{blue}{\textbf{3.48}} \\ \hline
U: Inpainting\_DAU & \underline{27.60}&\underline{24.19}&\underline{21.13}&\underline{18.47}&\underline{15.83}&\underline{13.48}&\underline{11.13}&\underline{9.64}&\underline{7.49} \\
U: \textbf{GLASS (unsupervised)} & \textbf{14.11}&\textbf{11.73}&\textbf{10.03}&\textbf{9.22}&\textbf{8.51}&\textbf{7.86}&\textbf{7.47}&\textbf{7.01}&\textbf{6.74} \\
\hline
\multicolumn{10}{c}{(c) New York subway (RMSE $/100$)} \\ \hline
B: OGTR & 9.63&4.88&3.38&2.87&2.59&2.37&2.16&1.98&1.77 \\
B: TRSS & 6.06&\textcolor{blue}{\underline{3.34}}&\textcolor{blue}{\underline{2.72}}&\textcolor{blue}{\underline{2.43}}&\textcolor{blue}{\underline{2.22}}&\textcolor{blue}{\underline{2.06}}&\textcolor{blue}{\underline{1.89}}&\textcolor{blue}{\underline{1.77}}&\textcolor{red}{\textbf{1.65}} \\
B: Tikhonov & \textcolor{blue}{\underline{5.23}}&3.78&3.31&3.01&2.78&2.57&2.38&2.21&2.04 \\
B: \textbf{GLASS (batch)} & \textcolor{red}{\textbf{3.61}}&\textcolor{red}{\textbf{2.76}}&\textcolor{red}{\textbf{2.43}}&\textcolor{red}{\textbf{2.23}}&\textcolor{red}{\textbf{2.09}}&\textcolor{red}{\textbf{1.97}}&\textcolor{red}{\textbf{1.85}}&\textcolor{red}{\textbf{1.76}}&\textcolor{blue}{\underline{1.66}} \\ \hline
S: GCN & 18.13&16.54&15.16&13.92&12.68&11.49&10.20&8.83&7.04 \\
S: ChebNet & \underline{16.88}&\underline{14.31}&\textbf{12.69}&\textbf{11.45}&\textbf{10.35}&\textbf{9.32}&\textbf{8.20}&\textbf{6.99}&\textbf{5.44} \\
S: LR & \textbf{15.14}&\textbf{13.89}&\underline{12.89}&\underline{11.96}&\underline{10.95}&\underline{9.91}&\underline{8.71}&\underline{7.31}&\underline{5.52} \\ \hline
O: OGTR (online) & \underline{8.51}&\underline{5.60}&\underline{4.56}&\underline{3.98}&\underline{3.45}&\underline{3.03}&\underline{2.67}&\underline{2.38}&\underline{2.01} \\
O: \textbf{GLASS (online)} & \textbf{6.46}&\textbf{4.40}&\textbf{3.60}&\textbf{3.15}&\textbf{2.75}&\textbf{2.44}&\textbf{2.19}&\textbf{2.00}&\textbf{1.80} \\ \hline
U: Inpainting\_DAU & \underline{22.43}&\underline{19.96}&\underline{18.41}&\underline{15.77}&\underline{14.07}&\underline{11.88}&\underline{10.05}&\underline{7.96}&\underline{5.66} \\
U: \textbf{GLASS (unsupervised)} & \textbf{5.71}&\textbf{4.35}&\textbf{3.72}&\textbf{3.30}&\textbf{3.01}&\textbf{2.76}&\textbf{2.51}&\textbf{2.35}&\textbf{2.13} \\
\hline
\end{tabular}

\end{table*}

\subsection{Datasets}
\subsubsection{Synthetic Dataset}
The synthetic dataset is created in the same way as in \cite{qiu2017TimeVarying}.
Within a square area of $100\times100$, 100 points are randomly selected according to a uniform distribution. 
To characterize the geographic adjacencies, a graph $\mathcal{G}$ is created by the $k$-nearest neighbor ($k$NN) method based on the Euclidean distances between the coordinates of the created points, and we set $k=5$. 
The weight of each edge is set to the reciprocal of the squared Euclidean distance between the connected nodes. All weights are then normalized by the maximum weight.
Based on the created graph $\mathcal{G}$, we generate a time-varying graph signal with $T=600$.
The initial low-frequency graph signal $\mathbf{x}_0$ at $t=0$ is defined as
\begin{equation}
\mathbf{x}_0 = \frac{S}{\|\mathbf{z}\|_2}\,\mathbf{z}, \quad \mathbf{z} = \mathbf{U}\,g(\mathbf{\Lambda})\,\mathbf{U}^\top \mathbf{f}_0,
\end{equation}
where $\mathbf{f}_0$ is a random vector whose elements follow $\mathcal{N}(0,1)$. The graph low-pass filter $g(\cdot)$ keeps the lowest 10\% of the graph frequencies and attenuates the others, i.e., $g(\lambda_i) = 1$ for $i < 0.1N$ and $g(\lambda_i) = 1/100$ otherwise. The constant $S=100$ scales the signal so that $\|\mathbf{x}_0\|_2 ^2 = 10^4$.
Using the generated $\mathbf{x}_0$, time-varying graph signal $\mathbf{x}_1,\dots,\mathbf{x}_{T-1}$ is recursively generated by
\begin{equation}
\mathbf{x}_t = \mathbf{x}_{t-1} + \mathbf{L}^{-1/2}\frac{\mathbf{f}_t}{\|\mathbf{f}_t\|_2},
\end{equation}
where $\mathbf{f}_t$ represents a signal variation at time $t$ and is a random vector whose elements follow $\mathcal{N}(0,1)$, and $\mathbf{L}^{-1/2} = \mathbf{U}\,\mathbf{\Lambda}^{-1/2}\,\mathbf{U}^\top$ with $\mathbf{\Lambda}^{-1/2} = \operatorname{diag}(0,\lambda_1^{-1/2},\dots,\lambda_{N-1}^{-1/2})$.
We generate 15 time-varying graph signals by the above method, of which 10 are used as the training set and 5 as the test set.

\subsubsection{Japan COVID-19 Number of Infected Patient Dataset}
We use the COVID-19 dataset published by the Japanese Ministry of Health, Labour and Welfare (MHLW)\footnote{The dataset is available at \url{https://covid19.mhlw.go.jp/public/opendata/newly_confirmed_cases_daily.csv}}.
This dataset records the daily number of newly infected patients in each prefecture of Japan ($N=47$) from Jan. 16, 2020 to May 8, 2023.
In the experiments, the data from Apr. 19, 2021 to Apr. 3, 2022 ($T=350$) are used as the training data, and those from May 24, 2022 to May 8, 2023 ($T=350$) as the test data.
The graph $\mathcal{G}$ is determined by the $k$-NN method ($k=5$) based on the 2-D coordinate of each prefecture.

\subsubsection{New York Subway Ridership Dataset}
We also use the NYC Subway Traffic Dataset\footnote{The dataset is available at \url{https://www.kaggle.com/datasets/eddeng/nyc-subway-traffic-data-20172021}}, which records the ridership of the New York City subway from 2017 to 2021.
In this dataset, the number of riders, i.e., the sum of the entries and the exits, is recorded for each station and each hour.
Stations are sometimes registered under several identifiers, and we first merge such duplicated entries. We then average the hourly counts over each day and exclude weekends, which gives one value per station per weekday.

In our experiments, we use all stations ($N=420$).
The graph is constructed from the subway map published by the MTA\footnote{The map is available at \url{https://new.mta.info/map/5256}}: two stations are connected by an edge of weight one if a subway line runs directly between them.
The training data is collected from Feb. 6, 2017 to Aug. 17, 2018 ($T=400$), and the test data is from Sep. 16, 2019 to Mar. 26, 2021 ($T=400$).
The 12 missing values in these periods are linearly interpolated along the time direction at each station.

\subsection{Restoration Results}

Table \ref{tab:ratio} shows the average RMSEs.
As shown in the table, GLASS (batch) achieves the lowest RMSEs in almost all cases.
Its advantage is largest when the observations are sparse.
This is because GLASS learns both the spatial and the temporal characteristics of the signal.

The static methods are far behind the batch and online methods on all the datasets.
They restore each time step independently, so they cannot exploit the temporal correlation of the signals.
The two graph neural networks do not consistently improve on the model-based LR, which suggests that the training data are not sufficient for them.

In the online setting, GLASS (online) outperforms OGTR (online) for every $\psi$ on the three datasets.
The difference comes from the multi-tap FIR filter and the regularization parameters, which GLASS learns in every layer.
GLASS (online) is also competitive with some of the existing batch methods.
On the synthetic dataset, it attains lower RMSEs than OGTR (batch) for every $\psi$.
On the Japan COVID-19 dataset, it outperforms OGTR (batch) at almost all $\psi$, Tikhonov for $\psi \geq 0.3$, and TRSS at $\psi=0.9$.
On the New York subway dataset, it also outperforms Tikhonov for $\psi \geq 0.5$.

GLASS (unsupervised) attains lower RMSEs than Inpainting DAU for every $\psi$ on the three datasets.
It is also competitive with the model-based batch methods when the observations are sparse.
At $\psi=0.1$, it outperforms TRSS and Tikhonov on the synthetic dataset, and TRSS on the New York subway dataset.
Even when the observations become dense, the gap remains small on these two datasets.

These results show that the three proposed methods are effective in every setting we consider.
GLASS (batch) is the most accurate method when the whole signal is available.
GLASS (online) keeps most of that accuracy, which makes it suited to real-time applications.
GLASS (unsupervised) stays close to the supervised batch methods without the ground truth of the target data.
The spatiotemporal regularizer is thus effective under the different constraints of the three settings.

\subsection{Selection of Hyperparameter $L$} \label{sec:exp-L}

\begin{figure}[t]
    \centering
    \includegraphics[width=8.5cm]{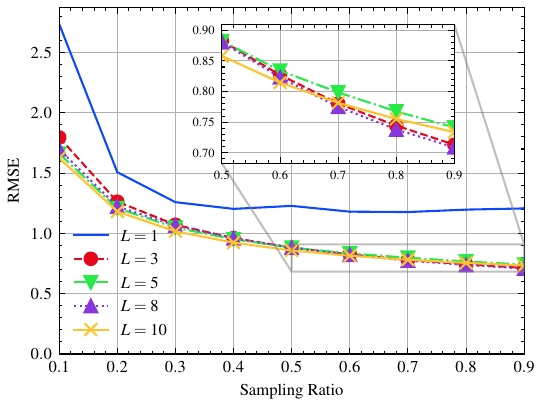}
    \caption{RMSE of GLASS (batch) on the synthetic dataset for several numbers of learnable taps $L$. The inset magnifies the region $\psi \geq 0.5$, where the curves for $L \geq 3$ are close to each other.}
    \label{fig:exp:each-L}
\end{figure}

The number of learnable taps $L$ is a hyperparameter of the proposed methods.
We therefore train GLASS (batch) on the synthetic dataset for $L \in \{1,3,5,8,10\}$ and compare the restoration accuracy.
All the other settings are the same as in the experiments above.

Figure \ref{fig:exp:each-L} shows the resulting RMSEs.
$L=1$ gives the largest RMSEs.
The normalization $\sum_{l=1}^{L} d_l = 1$ forces $d_1=1$ in this case, so the temporal filter cannot be learned.
The RMSE drops sharply at $L=3$, which shows that learning the FIR filter is effective.
Increasing $L$ further improves the accuracy at low sampling ratios, since the filter uses more past frames.
In the online setting, however, a larger $L$ makes the window longer and increases the computation per frame.
We therefore use $L=8$ in the other experiments as a compromise between the accuracy and the computational cost.

\begin{figure*}[tbp]
    \centering
    \subfigure[]{\includegraphics[width=5.7cm]{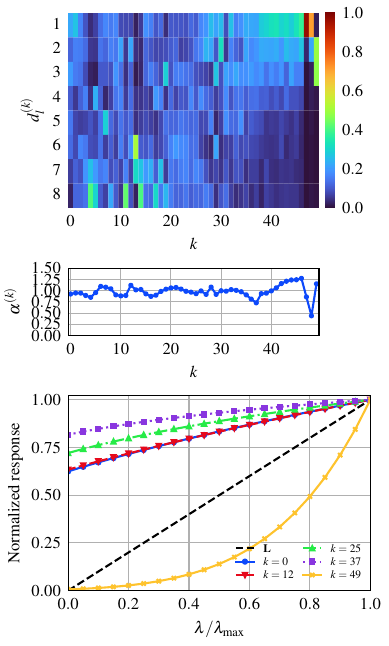}}%
    \subfigure[]{\includegraphics[width=5.7cm]{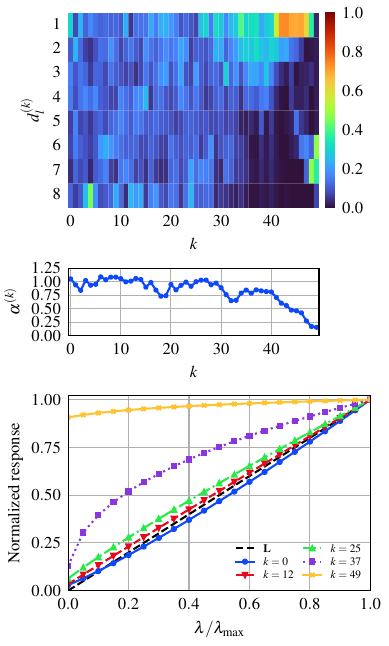}}%
    \subfigure[]{\includegraphics[width=5.7cm]{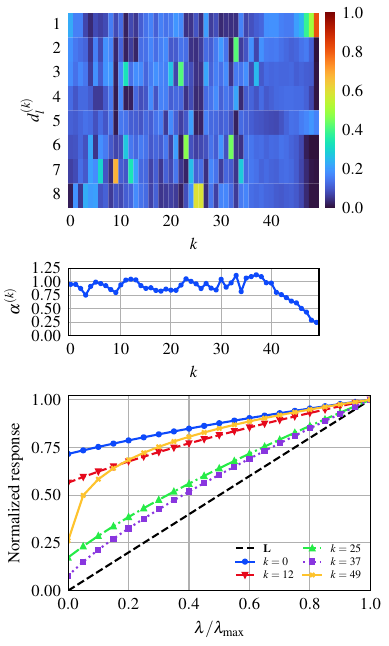}}
    \caption{Trained parameters of GLASS (batch) on (a) the synthetic, (b) the Japan COVID-19, and (c) the New York subway datasets. Top: the FIR filter coefficients $d_l^{(k)}$ of each layer. Middle: the regularization parameter $\alpha^{(k)}$. Bottom: the spectral response $(\lambda+\epsilon^{(k)})^{\beta^{(k)}}$ of representative layers, normalized so that it is one at $\lambda_{\max}$. The dashed line is the response $\lambda$ of the graph Laplacian.}
    \label{fig:exp:trained-params}
\end{figure*}

\begin{figure*}[tbp]
    \centering
    \vpanel{white}{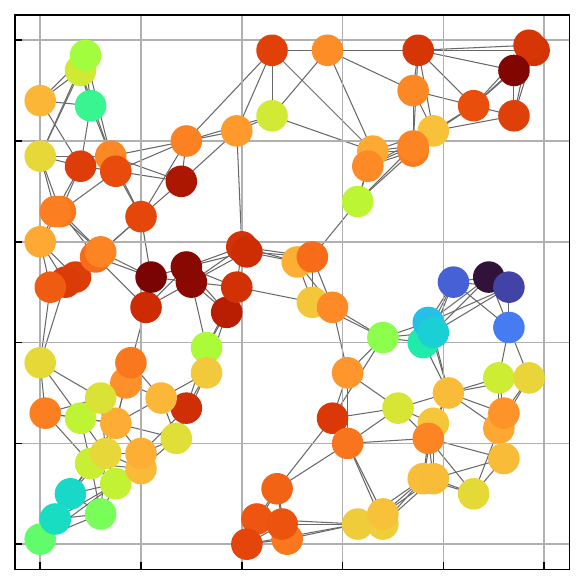}
    \vpanel{white}{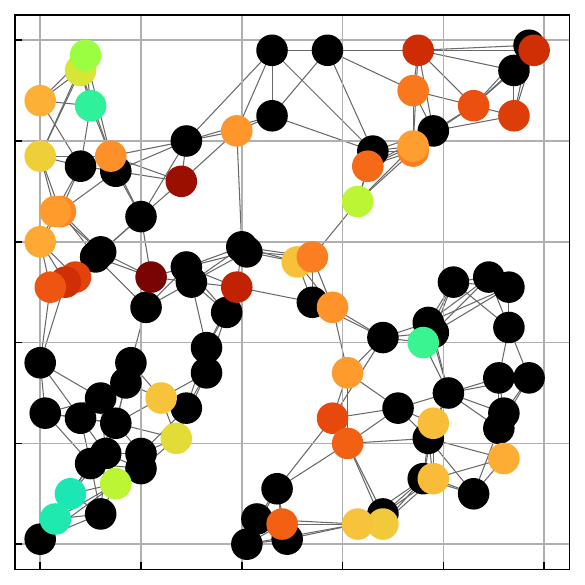}
    \vpanel{cyan}{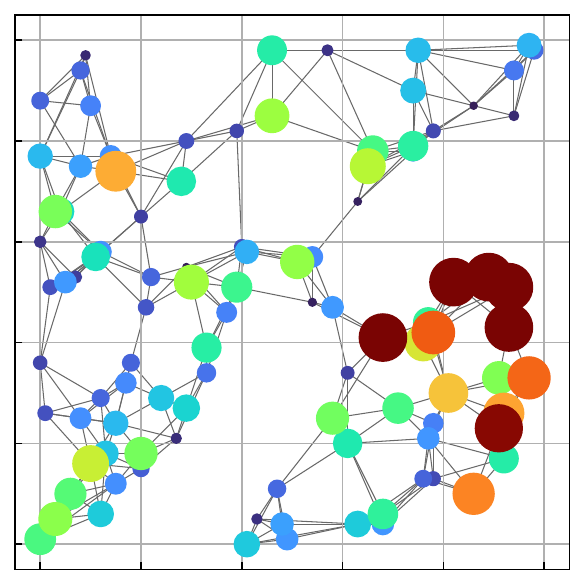}
    \vpanel{cyan}{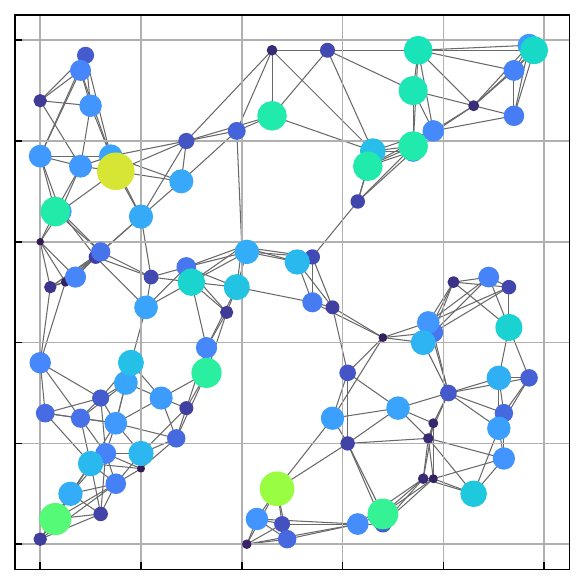}
    \vpanel{cyan}{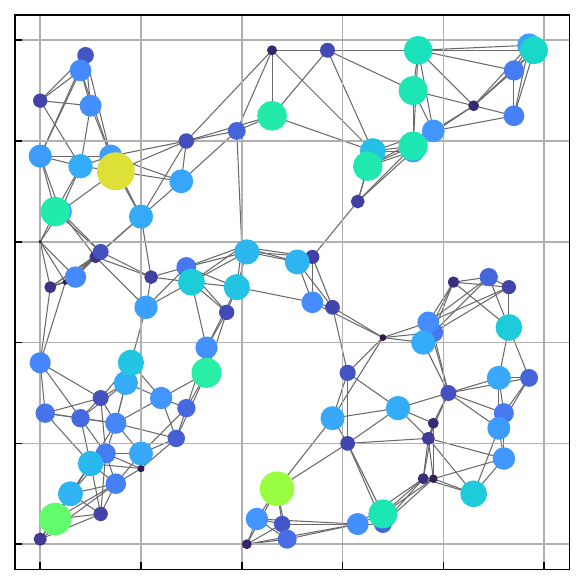}
    \\
    \vpanel{cyan}{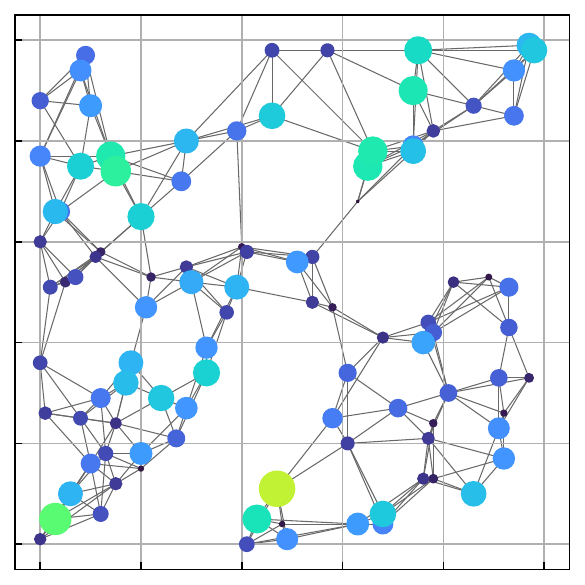}
    \vpanel{orange}{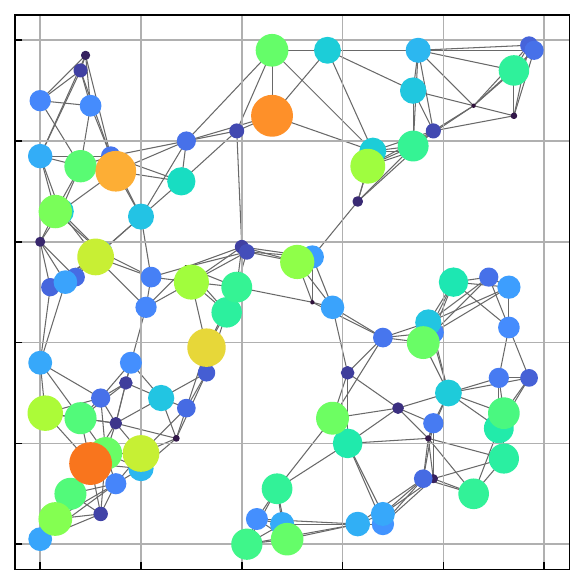}
    \vpanel{orange}{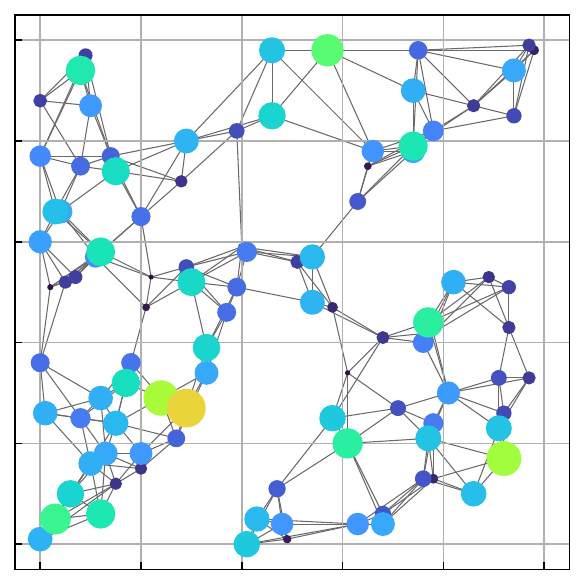}
    \vpanel{pink}{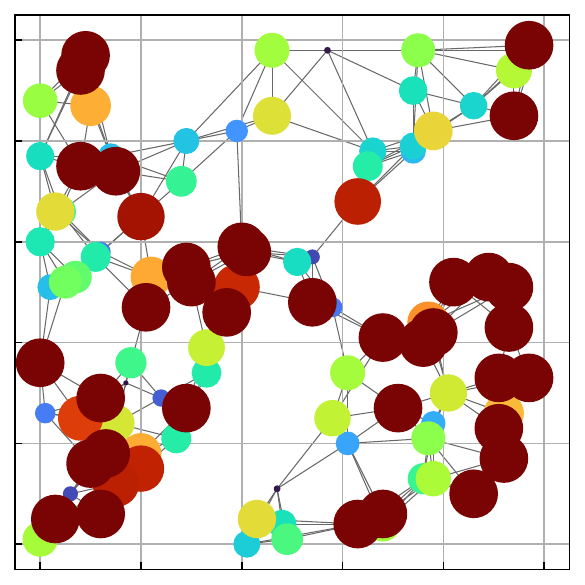}
    \vpanel{pink}{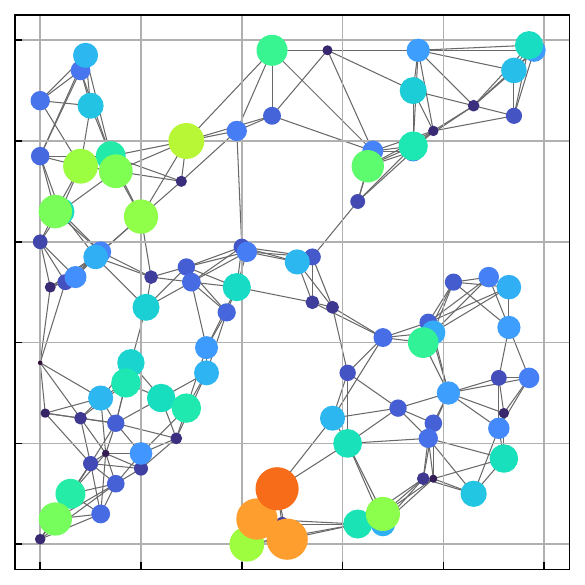}
    \\[1mm]
    (a), (b): \raisebox{-1mm}{\includegraphics[width=3.2cm]{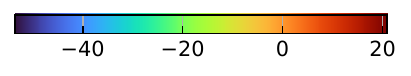}}\hspace{10mm}(c)--(j): \raisebox{-1mm}{\includegraphics[width=3.2cm]{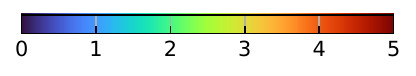}}
    \caption{Restoration results on the synthetic dataset ($t=350$, $\psi=0.5$, $\eta=0.1$). The panel colors cyan, orange, and pink indicate the batch, online, and unsupervised methods, respectively. (a) Ground truth. (b) Observed (RMSE: 12.88). (c) OGTR (3.33). (d) TRSS (1.01). (e) Tikhonov (1.01). (f) GLASS (batch) (0.92). (g) OGTR (online) (1.51). (h) GLASS (online) (1.08). (i) Inpainting DAU (9.93). (j) GLASS (unsupervised) (1.32). The black nodes in (b) are the missing ones. In (c)--(j), the color and the size of the nodes show the absolute difference from the ground truth.}
    \label{fig:exp:visualization-syn}
\end{figure*}

\begin{figure*}[tbp]
    \centering
    \vpanel{white}{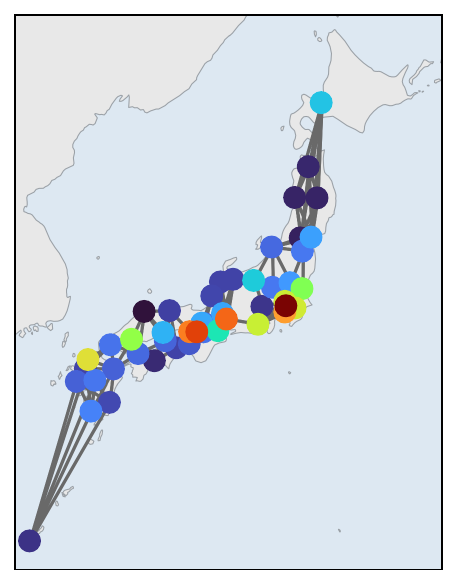}
    \vpanel{white}{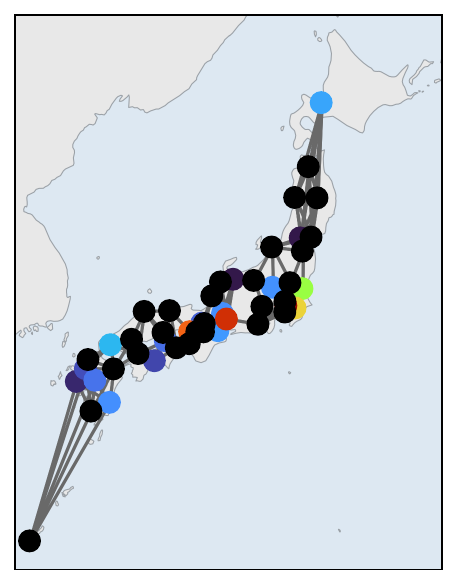}
    \vpanel{cyan}{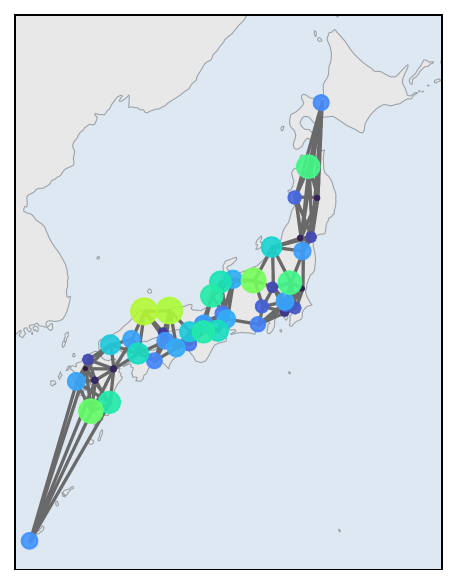}
    \vpanel{cyan}{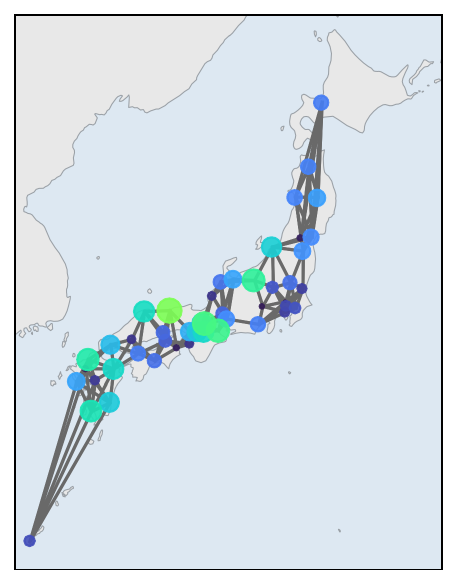}
    \vpanel{cyan}{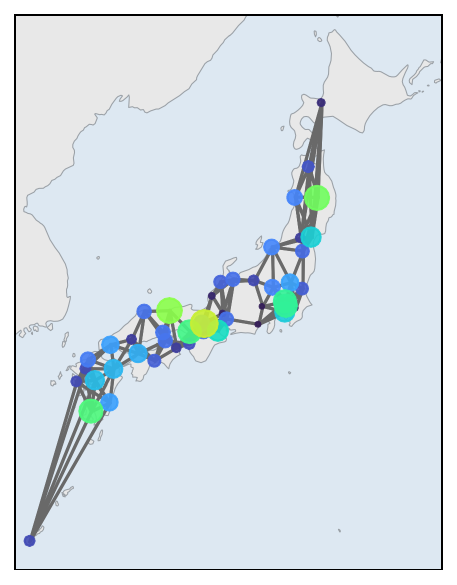}
    \\
    \vpanel{cyan}{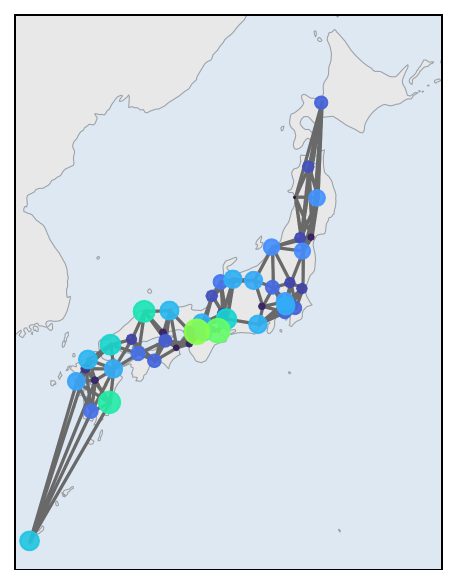}
    \vpanel{orange}{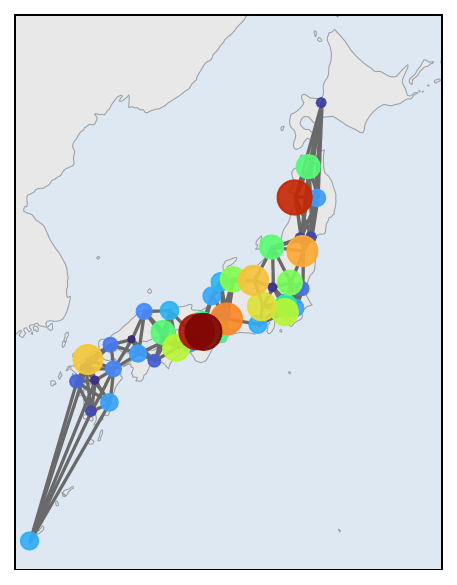}
    \vpanel{orange}{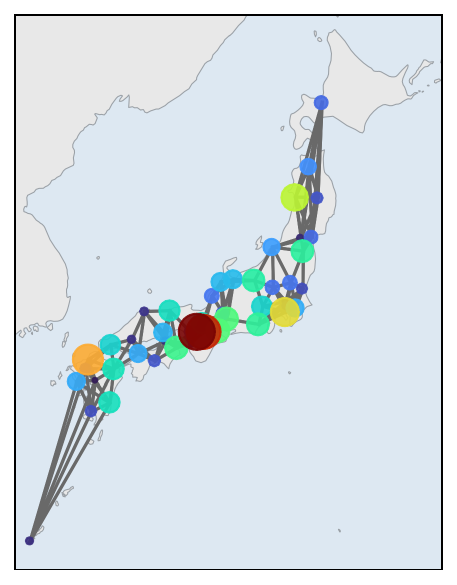}
    \vpanel{pink}{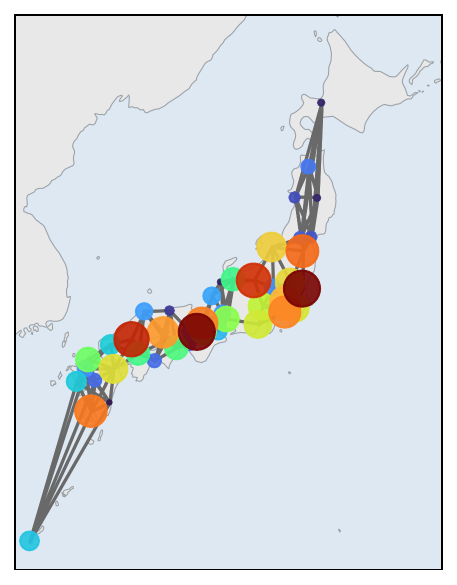}
    \vpanel{pink}{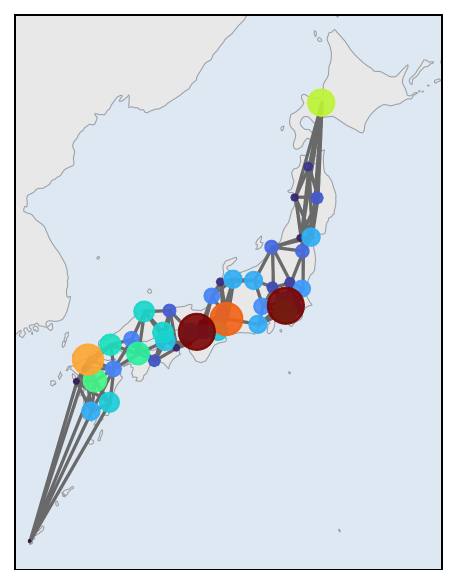}
    \\[1mm]
    (a), (b): \raisebox{-1mm}{\includegraphics[width=3.2cm]{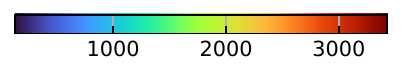}}\hspace{10mm}(c)--(j): \raisebox{-1mm}{\includegraphics[width=3.2cm]{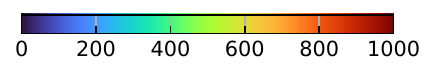}}
    \caption{Restoration results on the Japan COVID-19 dataset ($t=250$, $\psi=0.5$, $\eta=0.1$). The panel colors are the same as in Fig. \ref{fig:exp:visualization-syn}. (a) Ground truth. (b) Observed (RMSE: 887.13). (c) OGTR (246.12). (d) TRSS (212.51). (e) Tikhonov (220.02). (f) GLASS (batch) (193.84). (g) OGTR (online) (399.95). (h) GLASS (online) (368.44). (i) Inpainting DAU (506.39). (j) GLASS (unsupervised) (387.23). In (c)--(j), the color and the size of the nodes show the absolute difference from the ground truth.}
    \label{fig:exp:visualization-covid}
\end{figure*}

\begin{figure*}[tbp]
    \centering
    \vpanel{white}{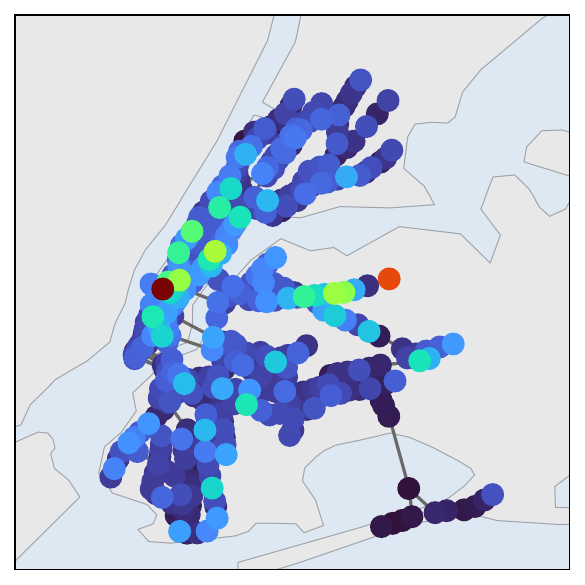}
    \vpanel{white}{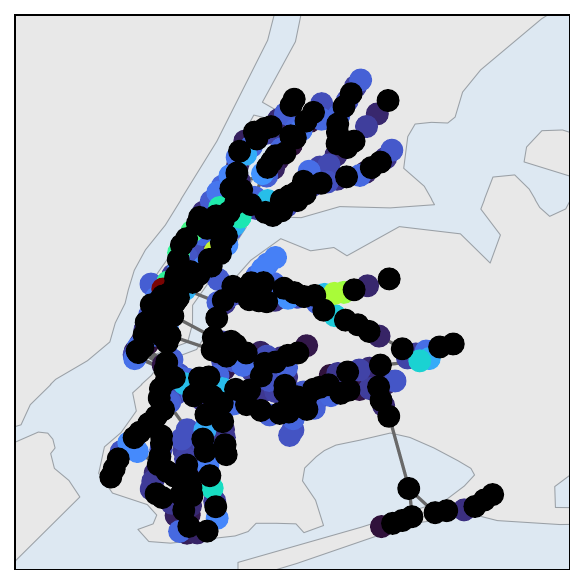}
    \vpanel{cyan}{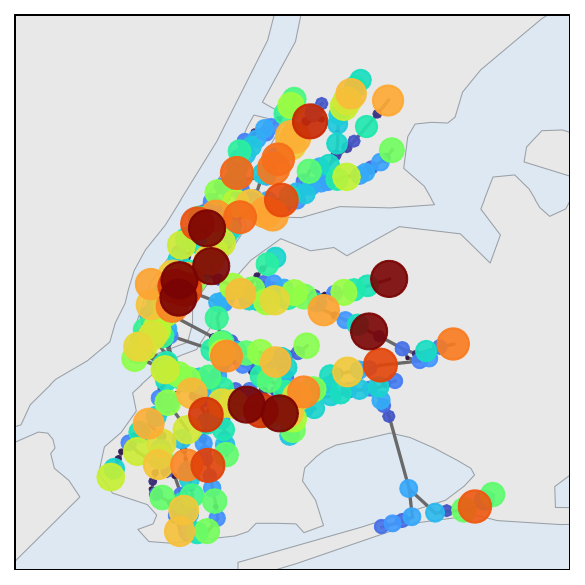}
    \vpanel{cyan}{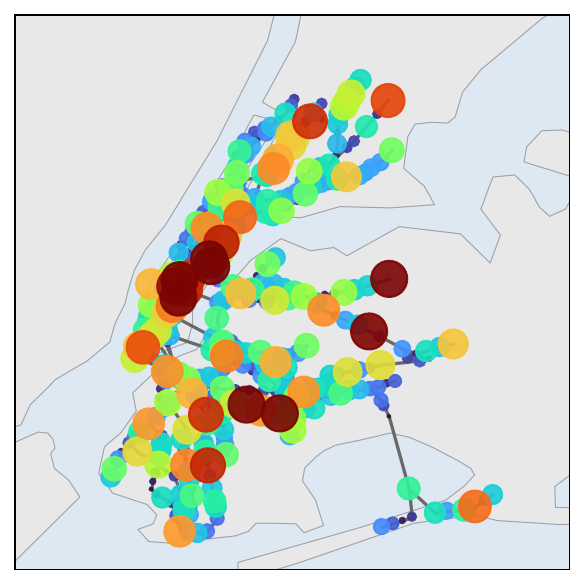}
    \vpanel{cyan}{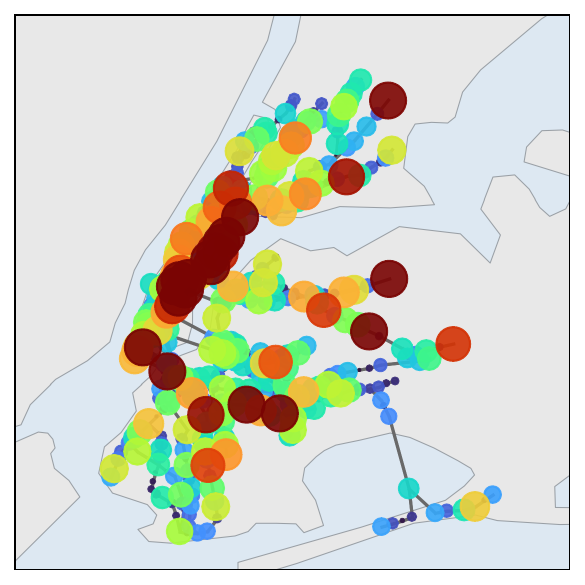}
    \\
    \vpanel{cyan}{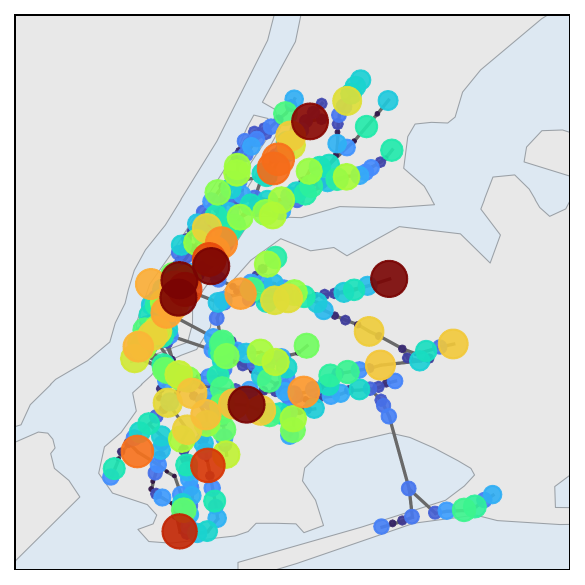}
    \vpanel{orange}{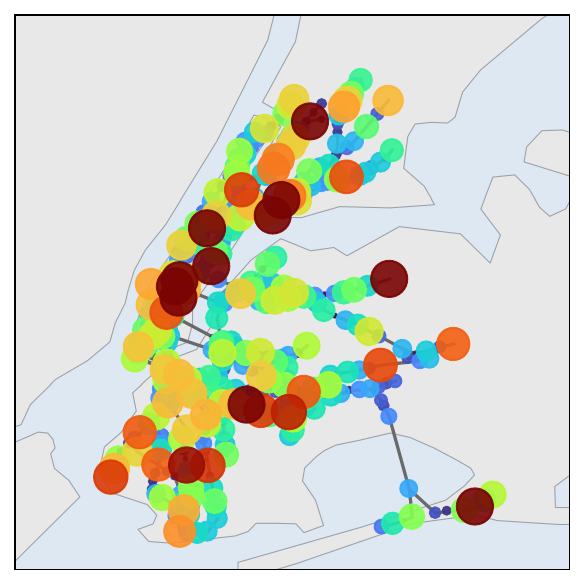}
    \vpanel{orange}{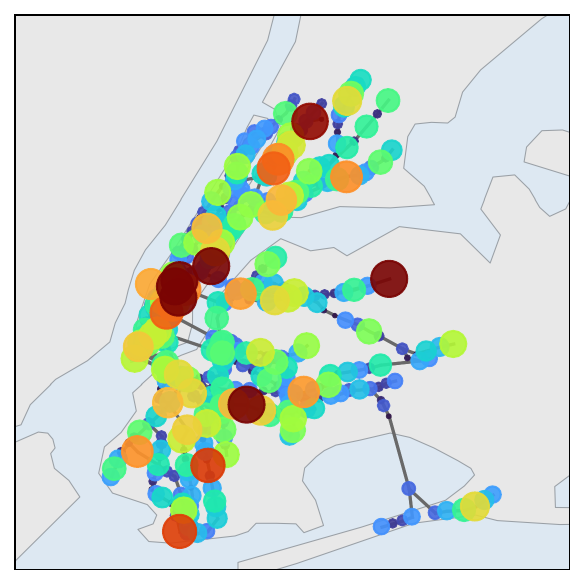}
    \vpanel{pink}{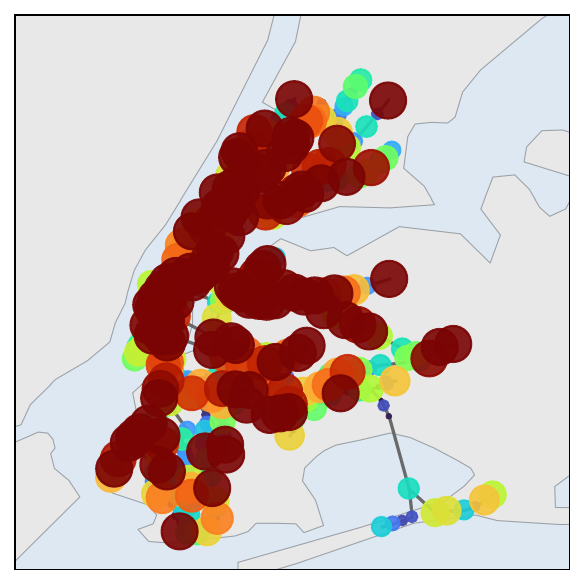}
    \vpanel{pink}{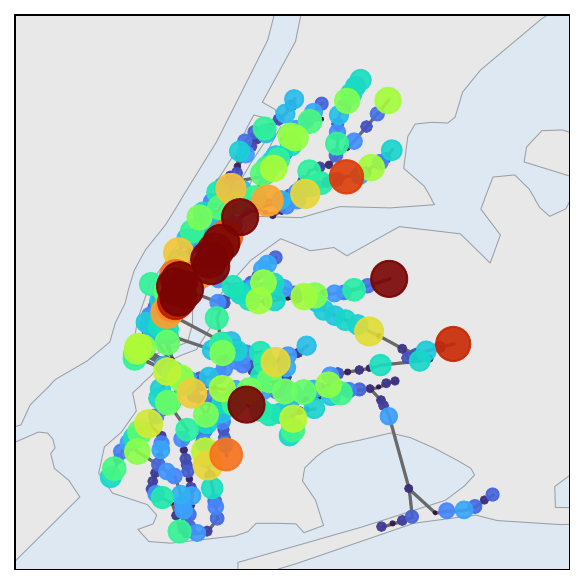}
    \\[1mm]
    (a), (b): \raisebox{-1mm}{\includegraphics[width=3.2cm]{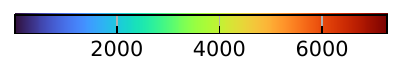}}\hspace{10mm}(c)--(j): \raisebox{-1mm}{\includegraphics[width=3.2cm]{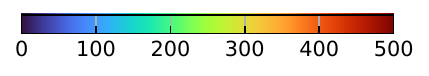}}
    \caption{Restoration results on the New York subway dataset ($t=350$, $\psi=0.5$, $\eta=0.1$). The panel colors are the same as in Fig. \ref{fig:exp:visualization-syn}. (a) Ground truth. (b) Observed (RMSE: 711.64). (c) OGTR (199.88). (d) TRSS (194.76). (e) Tikhonov (233.46). (f) GLASS (batch) (173.89). (g) OGTR (online) (212.39). (h) GLASS (online) (179.74). (i) Inpainting DAU (578.22). (j) GLASS (unsupervised) (178.95). In (c)--(j), the color and the size of the nodes show the absolute difference from the ground truth.}
    \label{fig:exp:visualization-nys}
\end{figure*}

\subsection{Discussion of Trained Parameters} \label{sec:exp-parameters}

Figure \ref{fig:exp:trained-params} shows the trained parameters of GLASS (batch) on the three datasets.
The FIR filter coefficients show a similar tendency on the three datasets.
The shallow layers spread the weight over the past frames, and the deep layers concentrate it on $d_1$.
The network thus uses the whole window first and refines the result from the most recent frame afterwards.
The regularization parameter $\alpha^{(k)}$ also decreases toward the last layers on the two real-world datasets, which weakens the regularization at the end of the network.

The learned spectral response is above the response $\lambda$ of the graph Laplacian in most layers.
The low graph frequencies are therefore penalized more than by the graph Laplacian.
The gap between the learned response and $\lambda$, however, differs across the datasets and the layers.
On the Japan COVID-19 dataset, it is raised the most in the deeper layers.
On the New York subway dataset, in contrast, it is raised the most in the shallow layers.
The spatial part of the regularizer is thus specific to each dataset, unlike the FIR filter coefficients.

\subsection{Visualization of Restored Signals} \label{sec:exp-visualization}

Figures \ref{fig:exp:visualization-syn}, \ref{fig:exp:visualization-covid}, and \ref{fig:exp:visualization-nys} show the restored signals of one time step on the three datasets.
We omit the static methods here, since their RMSEs are much larger than those of the other methods.

Figure \ref{fig:exp:visualization-syn} shows the results on the synthetic dataset.
Each proposed method keeps the error small over the whole graph in its own setting.
GLASS (batch) has smaller errors than OGTR, TRSS, and Tikhonov, and GLASS (online) has smaller errors than OGTR (online).
The difference is largest in the unsupervised setting.
For Inpainting DAU, a third of the nodes have an error larger than 5.
No node of GLASS (unsupervised) does.
The proposed methods are thus consistently accurate in the three settings.

Figure \ref{fig:exp:visualization-covid} shows the results on the Japan COVID-19 dataset, where the tendency is the same.
The nodes at which GLASS has a large error are the prefectures that contain a large city.
The number of cases is large there, and so is the absolute error.
Inpainting DAU has larger errors over the whole map, not only at these prefectures.

Figure \ref{fig:exp:visualization-nys} shows the results on the New York subway dataset.
The errors of every method concentrate on the stations with a large dynamic range.
Many passengers pass through these stations, and their signals change greatly over time.
GLASS keeps the error smaller than the existing methods even at these stations.

\section{Conclusion} \label{sec:conclusion}
In this paper, we propose supervised batch, supervised online, and unsupervised batch restoration methods for time-varying graph signals using deep algorithm unrolling (DAU).
The proposed method first formulates an optimization problem using a multi-tap FIR filter for temporal smoothness and the Sobolev operator for spatial smoothness.
The proposed supervised batch method learns the FIR filter coefficients and the regularization parameters from the whole observed signal. The proposed supervised online method learns them for a causal recursion, which restores each frame from a window of past restorations and the current observation.
The proposed unsupervised batch method achieves high-quality restoration for unknown data by learning them from the target observation alone.
In the experiments on several synthetic and real-world datasets, the proposed supervised batch method achieved the lowest RMSEs in almost all cases.
The proposed supervised online and unsupervised batch methods outperformed the existing methods of their own settings, and were often competitive with the existing batch methods.

\bibliographystyle{IEEEtran}
\bibliography{IEEEabrv,TVGSR-DAU_jounal}

\vfill

\end{document}